\documentclass[twocolumn, twocolappendix]{aastex631}

\usepackage{xspace}
\usepackage{amsmath}
\usepackage{amssymb}
\usepackage{physics}
\usepackage{multirow}

\newcommand{\SO}{SO\xspace}
\newcommand{\SOtwo}{SO$_2$\xspace}
\newcommand{\HtwoS}{H$_2$S\xspace}

\shorttitle{H$_2$S in HD 163296}
\shortauthors{Yamato et al.}
\graphicspath{{./}{figures/}}
\begin{document}
\title{Compact Hydrogen Sulfide Emission Indicates Sulfur-bearing Ice Sublimation in the Inner Disk of HD~163296}

\author[0000-0003-4099-6941]{Yoshihide Yamato}
\affiliation{RIKEN Pioneering Research Institute, 2-1 Hirosawa, Wako, Saitama 351-0198, Japan}
\affiliation{Department of Astronomy, Graduate School of Science, The University of Tokyo, 7-3-1 Hongo, Bunkyo, Tokyo 113-0033, Japan}
\email{yyamato.as@gmail.com}

\author[0000-0003-3283-6884]{Yuri Aikawa}
\affiliation{Department of Astronomy, Graduate School of Science, The University of Tokyo, 7-3-1 Hongo, Bunkyo, Tokyo 113-0033, Japan}

\author[0000-0002-2026-8157]{Kenji Furuya}
\affiliation{RIKEN Pioneering Research Institute, 2-1 Hirosawa, Wako, Saitama 351-0198, Japan}

\author[0000-0003-1413-1776]{Charles J.\ Law}
\altaffiliation{NASA Hubble Fellowship Program Sagan Fellow}
\affiliation{Department of Astronomy, University of Virginia, Charlottesville, VA 22904, USA}
% \email[show]{cjl8rd@virginia.edu}

\author[0000-0001-8798-1347]{Karin I. \"Oberg}
\affiliation{Center for Astrophysics | Harvard \& Smithsonian, 60 Garden St., Cambridge, MA 02138, USA}

\author[0000-0001-8642-1786]{Chunhua Qi}
\affiliation{Institute for Astrophysical Research, Boston University, 725 Commonwealth Avenue, Boston, MA 02215, USA}

\author[0000-0002-2700-9676]{Cataldi Gianni}
\affiliation{National Astronomical Observatory of Japan, Osawa 2-21-1, Mitaka, Tokyo 181-8588, Japan}

\author[0000-0003-1837-3772]{Romane Le Gal}
\affiliation{Universit\'e Grenoble Alpes, CNRS, IPAG, F-38000 Grenoble, France}
\affiliation{Institut de Radioastronomie Millimetrique (IRAM), 300 rue de la piscine, F-38406 Saint-Martin d’H`eres, France}

\author[0000-0003-2493-912X]{Shota Notsu}
\affiliation{Department of Earth and Planetary Science, Graduate School of Science, The University of Tokyo, 7-3-1 Hongo, Bunkyo, Tokyo 113-0033, Japan}
\affiliation{RIKEN Pioneering Research Institute, 2-1 Hirosawa, Wako, Saitama 351-0198, Japan}

\author[0000-0003-4784-3040]{Viviana V. Guzm\'an}
\affiliation{Instituto de Astrofísica, Pontificia Universidad Cat\'olica de Chile, Av. Vicu\~na Mackenna 4860, 7820436 Macul, Santiago, Chile}
\affiliation{$^{}$Millennium Nucleus on Young Exoplanets and their Moons (YEMS)}

\author[0000-0001-6947-6072]{Jane Huang}
\affiliation{Department of Astronomy, Columbia University, 538 W. 120th Street, Pupin Hall, New York, NY 10027, USA}



\begin{abstract}
The sulfur chemistry in protoplanetary disks directly affects the composition and potential habitability of nascent planets, but its volatile inventory remains highly uncertain. Here, we present deep Atacama Large Millimeter/submillimeter Array (ALMA) observations of hydrogen sulfide (H$_2$S) along with SO and SO$_2$ in the disk around HD~163296 at an angular resolution of $\approx0\farcs3$ (or $\approx30$\,au). We detect unresolved, compact emission of H$_2$S and SO (and tentatively SO$_2$) at the disk center with a broad line width of $\sim40$\,km s$^{-1}$, suggesting that the emission is originating from the innermost regions. By fitting line profiles with a geometrically-thin Keplerian-rotating disk model, we constrain the emitting radii and gas temperatures of these molecules to be $\approx$3--5\,au and $\gtrsim90$--120\,K, respectively, consistent with sublimation of sulfur-bearing molecules along with water ice in the inner warm region. While the higher or comparable column density of H$_2$S with respect to SO and SO$_2$ indicates that H$_2$S is an important volatile sulfur reservoir in the disk,
the limited constraints mean that we cannot rule out significantly depleted volatile sulfur as also commonly inferred in other planet-forming disks.
% within the current large uncertainties, the total amount of sulfur recovered does not depart from the scenario of depleted volatile sulfur in protoplanetary disks. 
Further observations are needed to better constrain disk sulfur inventory, unravel how sulfur compounds are reprocessed in disks, and shed light on the nature of less-volatile species, such as salts and sulfide minerals, which may occupy a significant portion of sulfur budget.         

\end{abstract}

%% \keywords{Classical Novae (251) --- Ultraviolet astronomy(1736) --- History of astronomy(1868) --- Interdisciplinary astronomy(804)}


\section{Introduction} \label{sec:intro}
Sulfur is tenth most abundant elements in the Universe (S/H $\approx1.3\times10^{-5}$; \citealt{Asplund2020}) and is essential for catalyzing life-related molecules and thus tightly linked to potential planet habitability \citep[e.g.,][]{Ranjan2018}. It is a well-known, long-standing problem that sulfur in dense star-forming gas is depleted by one to three orders of magnitude compared to the cosmic abundance \citep[e.g.,][]{Tieftruck1994, Fuente2023}. Infrared ice measurements toward molecular clouds and protostars, which only detect SO$_2$ and OCS and provide an upper limit on H$_2$S, also suggest that the icy sulfur compound accounts for only $\lesssim1$--5\% of the total sulfur budget \citep{Boogert1997, Boogert2015, McClure2023, Rocha2024}. 
% Recent experimental investigations on the ice absorption spectra obtained with the James Webb Space Telescope (JWST) suggest that ammonium hydrosulfide (NH$_4$SH) may be a potential sulfur carrier, perhaps accounting for up to 17--18\% of the total sulfur budget \citep{Slavicinska2025}. 
On the other hand, in-situ measurements by the Rosetta spacecraft in comet 67P/Churyummov-Gerasimenko (hereafter 67P/C-G) indicate that H$_2$S is the most abundant sulfur-bearing molecule 
% and that the total sulfur abundance is in agreement with the solar abundance, i.e., no depletion 
\citep{Calmonte2016}. Recent studies suggest that a semi-refractory species, ammonium hydrosulfide (NH$_4$$^+$SH$^-$) salt, can be a potential large sulfur carrier in comet 67P/C-G \citep{Altwegg2022}. While there is also a claim that NH$_4$$^+$SH$^-$ may occupy a significant portion of sulfur in the interstellar medium (ISM) as well \citep{Slavicinska2025}, the sulfur reservoir and its evolution during star and planet formation still remain an open question.

In protoplanetary disks, where planets and comets assemble, 
% and the disk chemical make-ups directly affect their composition,
sulfur is likely heavily depleted in the gas phase \citep{Keyte2024} and its majority ($\gtrsim90$\%) is thought to be locked into ices or refractory materials \citep{Kama2019}. Observations have detected a number of sulfur-bearing molecules in the outer region ($\gtrsim100$\,au) of disks, mainly CS and H$_2$CS (and their isotopologues; e.g., \citealt{Dutrey1997, Guilloteau2016, LeGal2019, Loomis2020, LeGal2021, Law2025, Law2025_CSsurvey, Teague2025}), which mainly traces the sulfur-depleted, cold gas. 
SO is also observed in a few Class II disks, which likely traces dynamical and/or planet-induced activities \citep{Law2023, Huang2023, Yoshida2024, Dutrey2024, Zagaria2025}, in addition to in younger disks where SO may trace shocks at the disk-envelope interface \citep[e.g.,][]{Sakai2014} and possibly ice sublimation in the innermost region \citep{Yamato2023}.
Recent sensitive observations with the Atacama Large Millimeter/submillimeter Array (ALMA) toward disks around Herbig Ae/Be stars have further provided a new insight into icy sulfur-bearing molecules in Class~II disks \citep[e.g.,][]{Booth2021, Booth2023}. For instance, \citet{Booth2021} detected bright SO and SO$_2$ (and their isotopologue) emission at the asymmetric dust trap in the Oph IRS~48 disk \citep[see also][]{Booth2024_IRS48, Temmink2025}. \citet{Booth2023} also detected SO emission in the inner region of the HD~100546 disk, potentially associated with an embedded protoplanet. Detection of these molecules in Herbig disks is thought to be predominantly an outcome of ice sublimation due to the higher disk temperature, as also supported by the detection of CH$_3$OH and other complex organics (e.g., \citealt{vanderMarel2021, Booth2021_HD100546, Brunken2022}). 
% These warm disks provide a way to indirectly observe icy molecules in disks.  

However, observations of (sublimated) H$_2$S ice in disks, the most dominant sulfur reservoir in comet 67P/C-G, remain scarce, and thus the budget of volatile sulfur (defined as sulfur-bearing molecules less- or equi-volatile to water) remains highly uncertain \citep[see e.g.,][]{Booth2024}. Gas-phase H$_2$S emission has been detected in a handful of disks \citep{Phuong2018, Riviere-Marichalar2021, Riviere-Marichalar2022} in a low-excitation transition with an upper state energy ($E_\mathrm{u}$) of $\approx28$\,K. The emission shows a ring-like distribution in the outer disk ($\gtrsim100$\,au), likely tracing cold H$_2$S gas due to the non-thermal desorption from dust grains \citep{Fuente2017, Oba2018}.
% but not directly thermally desorbed ices. 
\citet{Booth2025} detected weak emission of a high-excitation \HtwoS transition ($E_\mathrm{u}\approx168$\,K) at $\sim3\sigma$ significance in the HD~169142 disk, which is thus far the only indication of \HtwoS ice sublimation in the inner disk. Pure H$_2$S ice should have a (thermal) desorption temperature of $\sim60$--80\,K in the typical physical condition of a disk \citep{Collings2004, Santos2025}, while experimental studies suggest that a major fraction ($\sim75$--85\%) of H$_2$S ice is entrapped within H$_2$O ice, meaning that a majority of \HtwoS ice will likely sublimate along with water at a higher temperature \citep{Santos2025}. Constraining the sublimation front (i.e., snowline) of \HtwoS and its abundance by observations is crucial to understanding the sulfur chemistry and its evolution in protoplanetary disks.
% It is crucial to constrain the sublimation front (i.e., snowline) of H$_2$S and its abundance in protoplanetary disks by observations.

Here we present deep ALMA observations of H$_2$S, along with other sulfur-bearing molecules, SO and SO$_2$, in the HD~163296 disk. HD~163296 is a nearby ($d\approx 101$\,pc; \citealt{Gaia2023}) Herbig Ae star with a mass of $2.0\,M_\odot$ \citep{Teague2021}, a bolometric luminosity of $\sim17\,L_\odot$, and an age of $\gtrsim6$\,Myr \citep{Fairlamb2015}. Due to its proximity and large size, the disk around HD~163296 has extensively been studied in (sub-)mm wavelengths, including in both continuum and molecular line observations \citep[e.g.,][]{Isella2018, Oberg2021}. Multi-ringed structures in both dust continuum and multiple molecular line emission \citep{Isella2016, Isella2018, Guidi2022, Zhang2021, Ilee2021, Guzman2021}, as well as multiple kinematic evidence of embedded protoplanets, such as kinks, spirals, and meridional flows \citep{Pinte2018, Teague2019, Teague2021, Izuquierdo2022, Calcino2022}, have been identified. Molecular line observations have detected a large number of molecules mainly in the outer region of the disk \citep[e.g.,][]{Qi2013, Qi2015, Carney2017, Booth2019, Law2021, Guzman2021, Ilee2021, LeGal2021, Hernandez-Vera2024, Kashyap2025, Law2025}. This includes multi-line observations of CS, H$_2$CS, and their $^{34}$S isotopologues \citep{LeGal2021, Law2025}, where they found their multi-ringed structure in the outer region of the disk ($\gtrsim30\,$au) and potential $^{34}$S fractionation in both CS and H$_2$CS. However, the chemistry of the inner disk ($\lesssim10$\,au) is less explored compared to the outer disk due to the scarcity of molecular line observations sensitive to such scale.

This paper is organized as follows. We describe observational details in Section \ref{sec:observation} and analysis of the H$_2$S, SO, and SO$_2$ lines in Section \ref{sec:analysis}. We discuss the origin of the emission and sulfur reservoir in the HD~163296 disk and compare the result with other disks in Section \ref{sec:discussion}. We finally summarize the study in Section \ref{sec:summary}.
  
\section{Observation} \label{sec:observation}
We observed the HD~163296 disk in Band 7 during ALMA Cycle 8 (project code: 2021.1.00535.S, PI: Y, Yamato). Observations were carried out in four execution blocks. Table \ref{tab:observations} summarizes the observational details, including observation dates, number of antennas, on-source integration time, mean precepitable water vapor (PWV), baseline coverage, angular resolution, maximum recoverable scale (MRS), and calibrator information.

Initial calibration was performed by ALMA staff using the standard ALMA pipeline. Subsequent self-calibration and imaging were performed using the Common Astronomy Software Applications \citep[CASA;][]{CASA} modular version 6.6.4.34. The continuum visibilities were produced by averaging the line-free channels specified by visually inspecting the data cubes delivered from the observatory. The continuum emission peak on the image from each execution block was aligned to a common direction, $\alpha_\mathrm{ICRS} = 17^{\mathrm{h}}56^{\mathrm{m}}21\fs277, \delta_\mathrm{ICRS} = -21^{\mathrm{d}}57\arcmin22\farcs73$, which was used as the disk center in the following analysis. Then, seven rounds of phase-only self-calibration and one round of amplitude self-calibration were performed on the continuum data, resulting in a peak signal-to-noise ratio (S/N) increase by a factor of ${\approx}9$. The solutions were then applied to the non-averaged dataset. The continuum emission was finally subtracted from the line visibilities by fitting a first-order polynomial to line-free channels using the CASA task \texttt{uvcontsub}.

The SO $^3\Sigma\,J_N=7_7$--$6_6$ line ($E_\mathrm{u}=71$\,K) was covered by a dedicated spectral window with a resolution of 282 kHz ($\approx0.28$\,km\,s$^{-1}$), while ortho-H$_2$S $J_{K_a, K_c} = 3_{3,0}$--$3_{2,1}$ line ($E_\mathrm{u}=169$\,K) and two SO$_2$ lines were covered by the wide spectral window with a resolution of 1.1\,MHz ($\approx1.1$\,km\,s$^{-1}$) originally dedicated for continuum acquisition. The spectroscopic properties of these molecular lines are summarized in Table \ref{tab:lines}. 
% The continuum spectral window also covered a number of detected molecular lines including HC$_3$N, $c$-C$_3$H$_2$, H$_2$C$^{34}$S and CH$_2$CN, which will be presented in Appendix \ref{}.  
We imaged the visibilities of these spectral windows by \texttt{tclean} task in CASA with the modified Briggs' weighting scheme (\texttt{briggsbwtaper}) by cleaning down to 3$\times$ RMS levels. \textrm{We used \texttt{robust} $=0.5$ as we found it provides the highest S/Ns after testing different values (0.5, 1.0, and 2.0) for all three lines \textrm{(e.g., the peak S/Ns on the velocity-integrated intensity maps of the SO line are 5.5 and 4.9 for \texttt{robust} $=1.0$ and 2.0, respectively, both of which are lower than the S/N of the \texttt{robust} $=0.5$ map in Figure \ref{fig:mom0_gallery})}.} We generated image cubes with two different velocity channel widths of 1.2 km s$^{-1}$ and 4.8 km s$^{-1}$. The wider channel width intended to prioritize the weak line detections while still resolving the broad line width (see Section \ref{sec:analysis}). 
% We also generated another set of images with a \texttt{robust} parameter of 2.0 and a velocity channel width of 4.8 km s$^{-1}$, to explore any possible extended emission (see Appendix \ref{appendix:}). 
All the images were made using multiscale deconvolver with scales of [0, 5, 15, 25] pixels. The beam sizes and RMS noise levels of the image with \texttt{robust} $=0.5$ and a channel width of 1.2\,km\,s$^{-1}$ are reported in Table \ref{tab:lines}.

In addition to these lines, which are main focus of this study, the same dataset detects two other sulfur-bearing molecules, C$^{34}$S $J=6$--$5$ and ortho-H$_2$C$^{34}$S $J_{K_a, K_c}=9_{1,9}$--$8_{1,8}$, where the latter represents the first-ever detection of $^{34}$S isotopologues of H$_2$CS in a protoplantary disk. The detailed analysis combined with other archival data are shown in \citet{Law2025}. Furthermore, the continuum window serendipitously covers a few other molecular species detected, which is detailed in Appendix \ref{appendix:other_lines}.

% The continuum window also covers two SO$_2$ lines, which were not detected. The spectroscopic properties and observed flux densities of these molecular lines are summarized in Table \ref{tab:}

\begin{deluxetable*}{lcccccccc}
\label{tab:observations}
% \tablewidth{\textwidth}
\tablecaption{Observational Details}
\tablehead{\colhead{Date} & \colhead{\# of Ant.} & \colhead{On-source Int.} & \colhead{PWV}  & \colhead{Baseline} & \colhead{Ang. Res.} & \colhead{MRS} & \multicolumn{2}{c}{Calibrators}\\
\cline{8-9}
\colhead{} & \colhead{} & \colhead{(min)} & \colhead{(mm)} & \colhead{(m)} & \colhead{($\arcsec$)} & \colhead{($\arcsec$)} & \colhead{Bandpass/Amplitude} & \colhead{Phase}}
\startdata
2022 Jun. 9 & 41 & 43 & 0.7 & 15--783.5 & 0.3 & 4.2 & J1924-2914 & J1742-1517 \\
2022 Jun. 10 & 42 & 43 & 0.1 & 15--1213 & 0.3 & 4.1 & J1924-2914 & J1742-1517 \\
2022 Jun. 11 & 43 & 43 & 0.4 & 15--1213 & 0.3 & 4.3 & J1924-2914 & J1742-1517 \\
2022 Jun. 11 & 43 & 43 & 0.4 & 15--1213 & 0.3 & 4.3 & J1924-2914 & J1742-1517 
\enddata
\end{deluxetable*}

\begin{deluxetable*}{lccccccc}
\label{tab:lines}
% \tablewidth{\textwidth}
\tablecaption{Observed Molecular Lines}
\tablehead{\colhead{Transition} & \colhead{$\nu_0$} & \colhead{$E_\mathrm{u}$} & \colhead{$\log_{10}A_\mathrm{ul}$}  & \colhead{$g_\mathrm{u}$} & \colhead{Beam (P.A.)} & \colhead{RMS} & \colhead{$F_\nu$} \\
\colhead{} & \colhead{(GHz)} & \colhead{(K)} & \colhead{(s$^{-1}$)} & \colhead{} & \colhead{} & \colhead{(mJy beam$^{-1}$)} & \colhead{(mJy\,km\,s$^{-1}$)}} 
\startdata
ortho-H$_2$S $J_{K_a, K_c} = 3_{3,0}$--$3_{2,1}$ & 300.5055600 & 168.9 & $-$3.987 & 21 & $0\farcs37 \times0\farcs25\,(69\arcdeg)$ & 0.49 & $39\pm10$ \\
SO $3\Sigma\,J_N=7_7$--$6_6$ & 301.2861240 & 71.0 & $-$3.465 & 15 & $0\farcs37 \times0\farcs25\,(69\arcdeg)$ & 0.65 & $52\pm12$ \\
SO$_2$ $J_{K_a, K_c} = 19_{3,17}$--$19_{2,18}$ & 299.3168185 & 197.0 & $-$3.691 & 39 & $0\farcs37 \times0\farcs25\,(69\arcdeg)$ & 0.47 & $22\pm9$ \\
SO$_2$ $J_{K_a, K_c} = 32_{3,29}$--$32_{2,30}$ & 300.2734184 & 518.7 & $-$3.594  & 65 & $0\farcs37 \times0\farcs25\,(69\arcdeg)$ & 0.47 & $27\pm9$  
\enddata
\tablecomments{Beam size, RMS, and flux density are measured in images generated with \texttt{robust} $=0.5$ and channel widths of 1.2 km\,s$^{-1}$. The spectroscopic properties of the lines are taken from the Cologne Database for Molecular Spectroscopy (CDMS; \citealt{CDMS1, CDMS2, CDMS3}) with the original data from \citet{Helminger1972, Belov1995} for \HtwoS, from \citet{Clark1976, Klaus1996} for SO, and from \citet{Belov1998, Muller2005} for SO$_2$.}
\end{deluxetable*}

\section{Analysis \& results} \label{sec:analysis}
\begin{figure*}
\epsscale{1.15}
\plotone{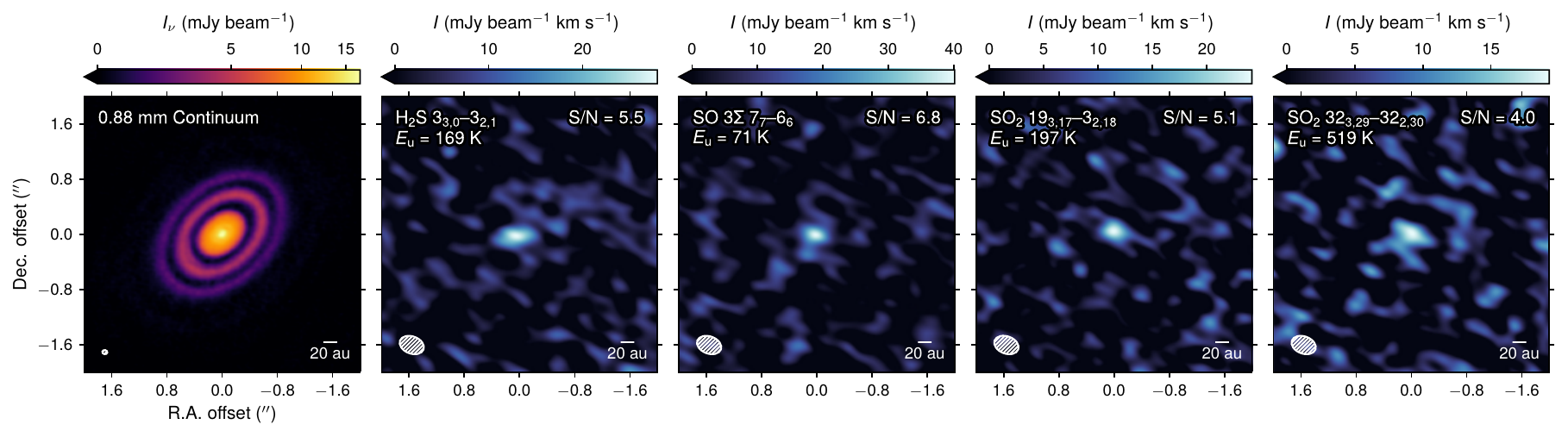}
\caption{High-resolution 0.88\,mm dust continuum image of the HD 163296 disk taken from \citet{Guidi2022} (left) and velocity-integrated intensity maps of \HtwoS, \SO, and two \SOtwo lines (others). The peak S/Ns of the emission are denoted in the upper right corner of each panel. For all panels, a 20\,au scale bar and the synthesized beam are indicated in the lower right and left corner, respectively. 
}
\label{fig:mom0_gallery}
\end{figure*}

\begin{figure*}
\epsscale{1.15}
\plotone{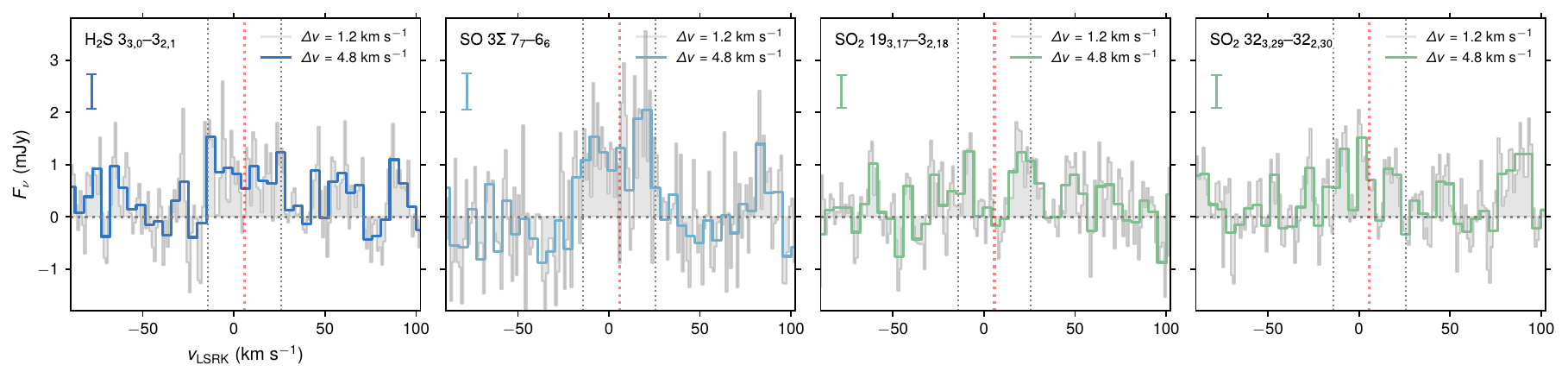}
\caption{Disk-integrated spectra of \HtwoS, \SO, and two \SOtwo lines. The gray and colored lines indicate the spectra at 1.2\,km\,s$^{-1}$ and 4.8\,km\,s$^{-1}$ channel widths, respectively. The vertical segment at the upper left corner indicate $1\sigma$ uncertainty of the spectra at 4.8\,km\,s$^{-1}$ channel width. \textrm{In each panel, the vertical red and gray dotted lines marks the systemic velocity of the source ($v_\mathrm{sys}=5.76$\,km\,s$^{-1}$) and the integration range for velocity-integrated intensity maps in Figure \ref{fig:mom0_gallery}, respectively.}
}
\label{fig:spectra}
\end{figure*}

Figure \ref{fig:mom0_gallery} shows the velocity-integrated intensity maps of the observed \SO, \HtwoS, and two \SOtwo lines along with the high-resolution 0.88\,mm dust continuum image taken from \citet{Guidi2022}. Using the Python package \texttt{bettermoments} \citep{bettermoments}, we spectrally integrated over $\pm20$\,km\,s$^{-1}$ with respect to the systemic velocity $v_\mathrm{sys}\approx5.76$\,km s$^{-1}$ \citep{Teague2019, Teague2021} for all lines with no intensity threshold for pixel inclusion. \textrm{This integration range encompasses all the apparent emission as shown in Figure \ref{fig:spectra}.} All maps exhibit spatially unresolved, compact emission at the disk center. The peak S/Ns of these lines are $\sim4$--7 as noted in each panel. Figure \ref{fig:spectra} shows the disk-integrated spectra of \HtwoS, \SO, and \SOtwo lines, extracted from the central two-beam region (i.e., an ellipse with major and minor axes of 2 $\times$ beam FWHM). \textrm{This extraction aperture ensures full flux recovery while avoiding unnecessary noise, as the emission is spatially unresolved}. The signals are quite weak with a narrower channel width of 1.2\,km\,s$^{-1}$, while they are detected in multiple consecutive channels at a channel width of 4.8\,km\,s$^{-1}$ for \HtwoS and \SO lines. Both \SO and \HtwoS spectra exhibit approximately symmetric line shape with respect to $v_\mathrm{sys}$, and a broad line width of ${\sim}40$\,km\,s$^{-1}$. For the two \SOtwo lines, the spectra lines do not show clear signals even with a wider channel width while showing peak S/Ns of $\sim4$--5 in the velocity-integrated intensity maps. We also performed spectral stacking analysis of the two \SOtwo lines, and found a $\approx3.6\sigma$ signal (see Appendix \ref{appendix:SO2_stacking}).
%, preventing us inferring robust detections.
The flux densities for these lines integrated over the same apertures as the spectrum extraction are reported in Table \ref{tab:lines}.

\begin{figure}
\epsscale{1.15}
\plotone{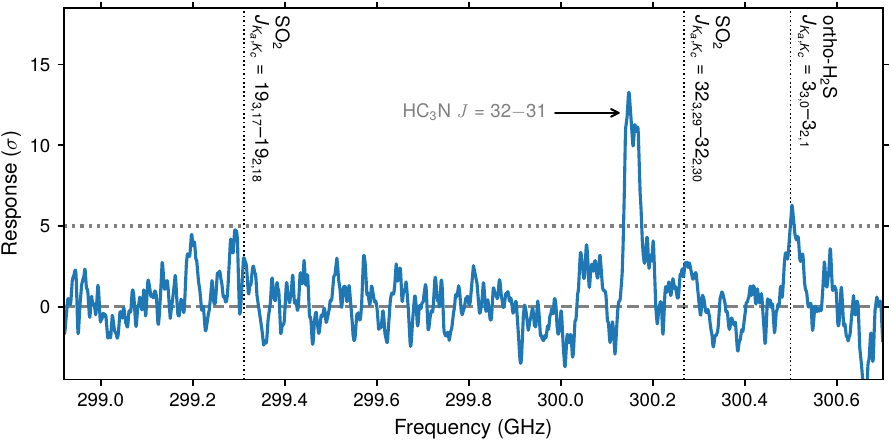}
\caption{Matched filter responses for the wide spectral window covering \HtwoS and \SOtwo lines. The vertical dotted lines mark the frequency of the targeted \HtwoS and \SOtwo lines. The horizontal dotted line indicates $5\sigma$ response, which is used to infer the detection. The strong feature at $\sim300.15$\,GHz is from the HC$_3$N $J=32$--$31$ line. 
}
\label{fig:matched_filter}
\end{figure}

To confirm the detection of these weak lines, we also performed a matched filter analysis \citep{Loomis2018}. Given that the emission shows a compact distribution and a broad line width, we used a simple Keplerian-rotating disk model with a small outer radius of 10\,au and known disk geometries ($\mathrm{P.A.}=133.3\arcdeg$, $i=46.7\arcdeg$; \citealt{Huang2018}) as a filter, which is cross-correlated with the observed visibilities to obtain the impulse responses. Figure \ref{fig:matched_filter} shows the responses for the wide spectral window covering \HtwoS and \SOtwo lines. While the \HtwoS line yields $\sim6$--$7\sigma$ response, the two \SOtwo lines show responses with $<5\sigma$. 
% We therefore claim a robust detection of \HtwoS (and \SO, which shows a clear feature on the image plane) and a tentative detection of \SOtwo. 
For \SO, the narrow spectral window compared to the line width (bandwidth / line width $\approx$ 3) prevented us from obtaining the accurate responses because the noise statistics are biased and the response becomes dominated by edge effects.

% for \HtwoS and \SOtwo lines. This analysis uses a template image cube as a filter, which is cross-correlated with the observed data on the visibility plane to obtain the response spectra. We used a simple Keplerian-rotating disk model with an outer radius of $3$\,au as a template for both \HtwoS and \SOtwo. For \SO, the narrow spectral window compared to the line width (bandwidth / line width $\approx$ 3--4) prevented us from obtaining the correct responses, as the . Figure \ref{fig:matched_filter} presents the resulting filter response for the wide spectral window covering \HtwoS and \SOtwo lines. While the \HtwoS line yields $\sim6$--$7\sigma$ response, the two \SOtwo lines show responses with $<5\sigma$. We therefore claim a robust detection of \HtwoS (and \SO, which shows a clear feature on the image plane) and a tentative detection of \SOtwo.
% This symmetric, broad line shape strongly indicates that the emission originates from the inner region of the Keplerian-rotating disk, in which the rotating velocity is larger at smaller radii.

\begin{figure*}
\epsscale{1.15}
\plotone{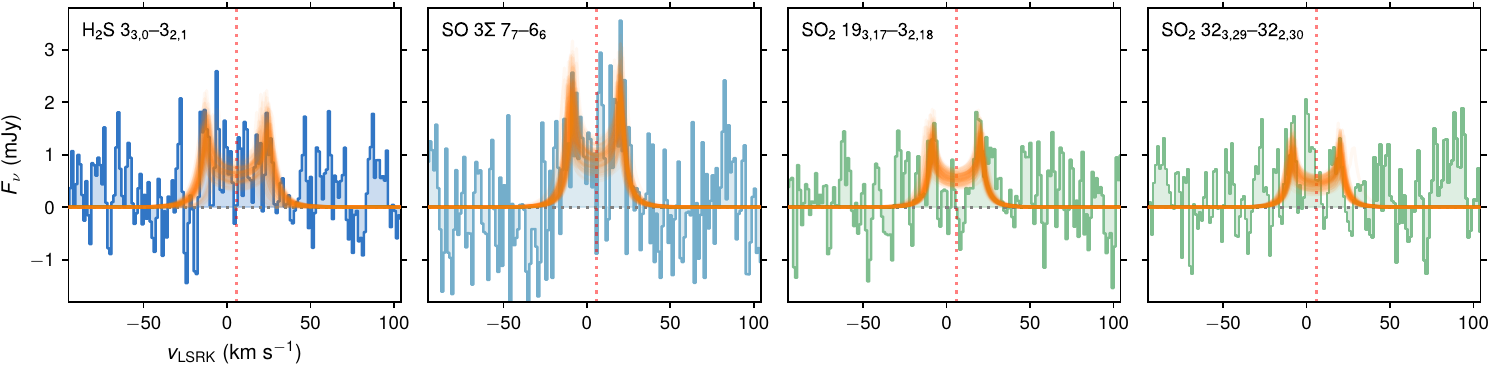}
\caption{Observed spectra of \HtwoS, \SO, and two \SOtwo lines at 1.2\,km\,s$^{-1}$ channel width overlaid with the fitted Keplerian disk model (orange). The model spectra are drawn for 100 samples randomly selected from the posterior chains. The vertical red dotted line in each panel marks the systemic velocity of the source.
}
\label{fig:spectra_fit}
\end{figure*}

\begin{figure*}
% \epsscale{1.15}
\plotone{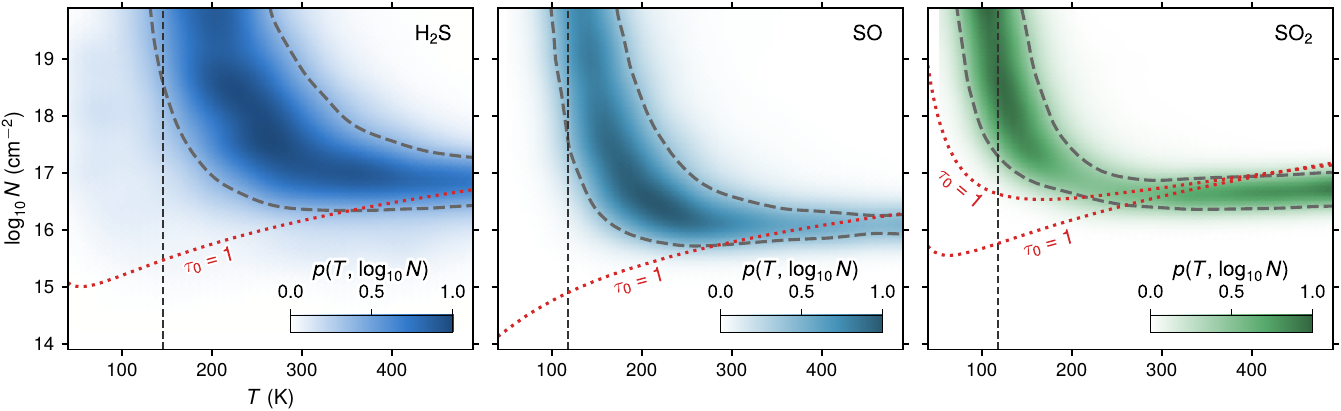}
\caption{Posterior distributions marginalized for temperature and column density for \HtwoS (left), \SO (middle), and \SOtwo (right). The color indicates the normalized probability distributions, while the gray dashed lines delineate 1$\sigma$ confidence interval. The red dotted lines mark the $\tau_0 = 1$ line for each transition. For \SOtwo, the upper line corresponds to the \SOtwo 32$_{3,29}$--32$_{2,30}$ transition, while the lower line is for the \SOtwo 19$_{3,17}$--19$_{2,18}$ transition. The region above these red dotted lines corresponds to the optically thick regime, while the region below them are optically thin regime. The vertical dashed lines mark the expected midplane temperature at the best-fit radius for each molecule based on the passively heated disk model (see Section \ref{subsubsec:sulfur_inner_outer}). The full covariances and marginalized distributions can be found in Figure \ref{fig:corner_plot} in Appendix \ref{appendix:corner_plot}.
}
\label{fig:N-T_plot}
\end{figure*}

While the spatially unresolved nature of the emission precludes us from directly inferring the emitting radii, the line profiles still provide information about the emitting radii under the assumption that the emitting gas follows Keplerian rotation. We model the observed line profiles (Figure \ref{fig:spectra}) by an axisymmetric, geometrically-thin Keplerian-rotating disk to quantitatively assess their emitting radii \citep[e.g.,][]{Bosman2021, Facchini2024}. The line of sight velocity of the emission from the location within a disk $(r,\phi)$ in the polar disk coordinate is 
\begin{equation}
    v_\mathrm{los} = v_\mathrm{Kep}\sin i \cos\phi + v_\mathrm{sys},
\end{equation}
where $v_\mathrm{Kep}=\sqrt{GM_\star/r}$ is the Keplerian velocity,  $G$ is the gravitational constant, $M_\star$ is the central stellar mass, and $i$ is the inclination of the disk. The line surface brightness at a given radius $r$ and azimuth $\phi$ is 
\begin{align}
    I_\nu(v; r, \phi) &= B_\nu(T) (1 - e^{-\tau(v)}) 
    % &\approx B_\nu(T)\tau(v),
\end{align}
where 
% the second line corresponds to the optically thin case, and 
\begin{equation}
    \tau(v) = \tau_0 \exp\left[-\frac{(v - v_\mathrm{los})^2}{2\sigma_v^2}\right],
\end{equation}
$T$ is the emitting gas temperature under the assumption of the local thermodynamical equilibrium (LTE) condition, $\tau_0$ is the optical depth at the line center, and $\sigma_v$ is the local line width. Assuming the LTE condition, the optical depth $\tau_0$ can be computed as 
\begin{equation}
    \tau_0 = \frac{c^3g_\mathrm{u}A_{\mathrm{ul}}N}{\sqrt{2\pi}\sigma_v\,8\pi\nu^3Q(T)}\left[\exp\left(\frac{h\nu}{k_\mathrm{B}T}\right) -1\right]\exp\left(-\frac{E_\mathrm{u}}{k_\mathrm{B}T}\right),
\end{equation}
where $N$ is the molecular column density, $Q(T)$ is the partition function at a temperature $T$, and $\nu, g_\mathrm{u}, A_\mathrm{ul}, E_\mathrm{u}$ are the line frequency, upper state degeneracy, Einstein coefficient for spontaneous emission, and upper state energy, respectively. 
% We \revise{assumed} a thermalized ortho-to-para ratio of 3, \revise{which is expected in the warm inner disk and also used in the previous observational studies of protostars \citep[e.g.,][]{Kushwahaa2023}}, to calculate total \HtwoS column density. 
\textrm{Given that thermally desorbed H$_2$O is expected to have a statistical ortho-to-para ratio \citep{Hama2018} and assuming that \HtwoS behaves similarly, we assumed} a statistical ortho-to-para ratio of 3 to calculate total \HtwoS column density. \textrm{This assumption is also used in the previous observational studies of protostars \citep[e.g.,][]{Kushwahaa2023}}.
The high density expected in the inner disk ($\gtrsim10^9$\,cm$^{-3}$; e.g.,  \citealt{Pirovano2022}) compared to the critical densities of the observed transitions ($\sim10^6$--$10^7$\,cm$^{-3}$)\footnote{Estimated based on the Einstein coefficients for spontaneous emission and the collisional rate coefficients available at the Leiden Atomic and Molecular Database (LAMDA; \citealt{LAMDA})} justifies the LTE assumption. We further assumed that the local line width is dominated by thermal broadening\footnote{\textrm{Note that thermal line widths at temperatures of 100--300\,K are 0.2--0.6 km s$^{-1}$ for three molecules, which are smaller than the channel width of the data (1.2\,km s$^{-1}$). This assumption thus does not significantly affect the emitting radius estimates unless that the gas temperature is extremely high ($\gtrsim1000$\,K) or that there are additional extreme line broadening (e.g., high turbulence with a Mach number of $\gg1$), which are both highly unlikely scenarios in a protoplanetary disk.}}, i.e., 
\begin{equation}
    \sigma_v = \sqrt{\frac{k_\mathrm{B}T}{\mu m_\mathrm{p}}},
\end{equation}
where $k_\mathrm{B}$ is the Boltzmann constant, $\mu$ is molecular weight, and $m_\mathrm{p}$ is the proton mass. The flux density of the lines integrated over the disk can then be numerically computed as 
\begin{equation}
    F_\nu(v) = \int_0^{2\pi}\int_{0}^{R}I_\nu(v; r, \phi)r\dd r\dd\phi,
\end{equation}
which was fitted to the observed line profile with an 1.2\,km\,s$^{-1}$ velocity channel width (Figure \ref{fig:spectra}). We left three parameters free for each molecular species: emitting radius $R$, gas temperature $T$, and column density $N$. For \SOtwo, which is only tentatively detected \textrm{and do not have sufficient S/Ns to independently constrain the emitting radius}, we assumed that the emitting radius $R$ is the same as that of SO based on their similar spatial distributions in other protoplanetary disks \citep[e.g.,][]{Semenov2018, Booth2024_HD100546, Booth2024_IRS48}. \textrm{Their actual distributions can differ depending on the specific physical condition of a disk, but we need a better sensitivity data to test this\footnote{At least the current model that assumed the same emitting radius as that of \SO reproduces well the stacked \SOtwo spectrum that have a slightly better S/N (see Figure \ref{fig:SO2_stacked} in Appendix \ref{appendix:SO2_stacking}).}.} We also left the systemic velocity $v_\mathrm{sys}$ as a common free parameter for all lines, resulting in a total of nine free parameters. \textrm{With these parameters, we constructed a likelihood function that is proportional to $\exp(-\chi^2/2)$, where $\chi^2 = \sum(F_{\nu,\mathrm{data}} - F_{\nu, \mathrm{model}})^2/\sigma^2$. We considered a flux calibration uncertainty of \textrm{10}\% in the likelihood function, in addition to the statistical noise.} The posterior probability distribution was explored using the affine-invariant Markov Chain Monte Carlo (MCMC) code \texttt{emcee} \citep{emcee} to generate posterior distributions of these parameters. We used uniform priors, i.e., $\mathcal{U}(1,40)$ for $R$ in au, $\mathcal{U}(20,500)$ for $T$ in K, $\mathcal{U}(14,20)$ for $\log_{10}N$ (cm$^{-2}$), and $\mathcal{U}(0, 11)$ for $v_\mathrm{sys}$ in km\,s$^{-1}$, where $\mathcal{U}(a,b)$ denotes the uniform distributions between $a$ and $b$. \textrm{These priors are common for all molecular species, and the ranges of priors are determined based on the spatial distributions of the emission (Figure \ref{fig:mom0_gallery}) for $R$, previous disk models of HD~163296 for $T$ and $\log_{10}N$ \citep{Pirovano2022}, and the known systemic velocity for $v_\mathrm{sys}$ \citep{Teague2019, Teague2021}.} \textrm{The temperature prior ranges are also consistent with the values inferred from the dust continuum and molecular line observations of the HD~163296 disk \citep[e.g.,][]{Calahan2021, Guidi2022}.} We used 100 walkers to sample in 15,000 steps, of which initial 2,000 steps are discarded as burn-in. Figure \ref{fig:spectra_fit} shows the fitted line profiles. 
% while the parameter covariances and marginalized posterior distributions are presented in Figure \ref{} in Appendix \ref{}. 
The characteristics of the observed spectra, including their broad line widths, are well reproduced with the model at least for \HtwoS and \SO lines that are robustly detected. The constraints on the parameters are reported in Table \ref{tab:lineprofilefits}. The emitting radii $R$ are $\sim3$--5\,au, consistent with the spatially unresolved emission at $\sim0\farcs3$ resolution (corresponding to $\sim30$\,au). As we observe only one or two transition(s) for each species, gas temperature $T$ and column densities $N$ are highly degenerate, and the solutions are mostly in the optically thick regime except for high temperatures (Figure \ref{fig:N-T_plot}). We calculated the lower end of 95\% highest density interval of the posterior samples as their lower limits (Table \ref{tab:lineprofilefits}). The systemic velocity $v_\mathrm{sys}$ derived here is consistent with the known literature value \citep[$\approx5.76$\,km\,s$^{-1}$; e.g.,][]{Teague2021}. 
The full marginalized posterior distributions and covariances between these parameters are shown in Figure \ref{fig:corner_plot} in Appendix \ref{appendix:corner_plot}. 

% Hence, the flux density for the emission originating from an area element $\dd\Omega = |r\dd r\dd\phi|$ is
% \begin{equation}
%     \dd F = I(r)\,\dd\Omega = I(r)\,r\dd r\frac{\dd v}{\sqrt{v_\mathrm{max}^2 - (v-v_\mathrm{sys})^2}},
% \end{equation}
% where $I(r)$ is the velocity-integrated intensity at radius $r$, and $v_\mathrm{max}:= v_\mathrm{Kep}\sin i$ (see Figure 11 in \citealt{Bosman2021} for the exact shape of this function). Therefore, the spectrum (i.e., line profile) integrated over $[r_\mathrm{in}, r_\mathrm{out}]$ can be written as 
% \begin{equation}
%     F(v) = \int_{r_\mathrm{in}}^{r_\mathrm{out}}\dv{F}{v} = \int_{r_\mathrm{in}}^{r_\mathrm{out}}\frac{I(r)r}{\sqrt{v_\mathrm{max}^2 - (v-v_\mathrm{sys})^2}}\dd r.
% \end{equation}
% While this formulation implicitly ignores the line broadening due to the local effect (e.g., thermal broadening), this should have a minimal effect given that the spectral resolution (${\approx}1.2$ km s$^{-1}$) is much coarser than the expected thermal line width ($\lesssim0.3$\,km\,s$^{-1}$ at $\lesssim300$\,K).

% We further parameterize $I(r)$ by molecular column density $N$ and excitation temperature $T_\mathrm{ex}$ as follows:
% \begin{equation}
%     I(r) = 
%     \begin{cases}
%         I & \quad r_\mathrm{in} \le r \le r_\mathrm{out}, \\
%         0 & \quad r_\mathrm{out} < r,
%     \end{cases}
% \end{equation}
% that is, 

\begin{deluxetable}{lcccc}
\label{tab:lineprofilefits}
% \tablewidth{\textwidth}
\tablecaption{Results of Line Profile Fits}
\tablehead{\colhead{Species} & \colhead{$R$} & \colhead{$T$} & \colhead{$N$} & \colhead{$v_\mathrm{sys}$} \\ 
\colhead{} & \colhead{(au)} & \colhead{(K)} & \colhead{(cm$^{-2}$)} & \colhead{(km\,s$^{-1}$)}
}
\startdata
H$_2$S & $3.0_{-0.3}^{+0.6}$ & $>120$ & $>1.0\times10^{16}$ & \multirow{3}{*}{$5.7^{+0.7}_{-0.9}$}\\
SO & \multirow{2}{*}{$4.6^{+0.5}_{-0.6}$} & $>92$ & $>5.7\times 10^{15}$ \\
SO$_2$ &  & $>85$ & $>2.1\times10^{16}$
\enddata
\tablecomments{The lower limits of $T$ and $N$ are defined as the lower end of 95\% highest density interval 
% (See also the posterior distributions in Figure \ref{})
.}
\end{deluxetable}

\section{Discussion} \label{sec:discussion}

\subsection{Origin of the Emission}
\subsubsection{Sublimation of Sulfur-rich Ice and Subsequent Gas-phase Processing}\label{subsubsec:S_ice_sublimation}
% A number of sulfur-bearing molecules has been detected in protoplanetary disks, including CS, H$_2$CS, CCS, \HtwoS, SO, and SO$_2$ \citep[and their isotopologues; e.g.,][]{Dutrey1997, Guilloteau2013, Semenov2018, LeGal2019, LeGal2021, Phuong2018, Phuong2021, Booth2023, Booth2024_IRS48}. Among them, 
\HtwoS has been detected in a handful of disks to date \citep{Phuong2018, Riviere-Marichalar2021, Riviere-Marichalar2022}, where these studies have observed a low-excitation \HtwoS line at 168.763\,GHz ($J_{K_a, K_c}=1_{1,0}$--$1_{0,1}$, $E_\mathrm{u}\approx28$\,K). The spatially resolved observations of this line toward the GG Tau disk and the AB Aur disk both show ring-like distributions at $\sim100$--$300$\,au radius and low column densities ($\sim10^{12}$--$10^{13}$\,cm$^{-2}$), suggesting that the observed \HtwoS emission traces the cold \HtwoS gas predominantly produced by non-thermal desorption \citep{Phuong2018, Riviere-Marichalar2022}. In contrast, the present observations detected a high-excitation \HtwoS line ($E_\mathrm{u} \approx 169$\,K), and its emission is concentrated on the central region of the disk. This is similar to the case of the disk around HD~169142, where the weak emission of the same \HtwoS transition has recently been detected \citep{Booth2025}. In the present work, we show that the emission traces the warm \HtwoS gas in the inner region by quantifying the emitting radius and gas temperature to be $\sim3$\,au and $>120$\,K. 

The emitting radius of \HtwoS coincides well with the radius at which the disk temperature reaches $\sim150$\,K in the midplane (or hypothetical water desorption front, $\sim2$--5\,au) constrained from \textit{Herschel} observations of the CO ladder \citep[$E_\mathrm{u}\approx61$--1,070\,K;][]{Pirovano2022}, and is in line with the constraints on the water snowline radius from sub-mm water line observations \citep[$\lesssim20$\,au;][]{Notsu2019}. The high \HtwoS gas temperature of $>120$\,K is also consistent with the typical sublimation temperature of water ($\sim150$\,K). These facts indicate that the observed emission traces the sublimation of H$_2$S ice along with water ice, which is fully consistent with experimental studies that suggest the majority ($\sim75$--$85$\%) of \HtwoS ice is entrapped within the water ice matrix and will co-desorb with water ice \citep{Santos2025}. 
% It is also possible that the observed emission traces the sublimation of NH$_4$$^+$SH$^-$, as it has a similar sublimation temperature to that of water \citep{Slavicinska2025} and readily dissociates into NH$_3$ and H$_2$S upon sublimation. It is challenging to distinguish the sublimation of neutral ice or NH$_4$$^+$SH$^-$, salt, with observations of sublimation products \citep{Slavicinska2025}.

% While the high temperature of $>120$\,K indicates that the \HtwoS emission detected traces the \HtwoS ice sublimated along with water, the present observations are likely not sensitive enough to detect the pure \HtwoS ice component.
% The latter component is likely indistinguishable from the abundant former component in the present observations due to the limited sensitivity.

% It is the detection of the warm \HtwoS gas that support the sublimation of sulfur-rich ice, since \HtwoS can form only on the dust grain surfaces while other sulfur-bearing molecules detected, namely, SO and SO$_2$, can form in the gas-phase as well. 

% While we have detected the warm \HtwoS gas for the first time in protoplanetary disks, \SO and \SOtwo (and their isotopologue) have previously been detected in several warm disks around Herbig Ae/Be stars as well \citep{Booth2023, Booth2023_HD169142, Booth2024_HD100546, Booth2024_IRS48}.  

% It is the detection of the warm \HtwoS gas that strongly suggest the sublimation of sulfur-rich ice (or salt).
% , of which the present observations show direct evidence for the first time in a protoplanetary disk. 
While \SO and \SOtwo have gas-phase formation routes in addition to the grain surface ones, \HtwoS mainly forms on grain surface via successive hydrogenation of S atoms and thus essentially points to the sublimation of sulfur-bearing ice. SO can efficiently form in the gas-phase via $\mathrm{SH} + \mathrm{O} \rightarrow \mathrm{SO} + \mathrm{H}$ or $\mathrm{S} + \mathrm{OH} \rightarrow \mathrm{SO} + \mathrm{H}$, where SH and OH are the photodissociation products of sublimated \HtwoS and H$_2$O, respectively \citep[e.g.,][]{Semenov2018}. \SOtwo is one of the detected sulfur-bearing ices in molecular clouds \citep[e.g.,][]{McClure2023}, but it can also form via $\mathrm{SO} + \mathrm{OH} \rightarrow \mathrm{SO_2} + \mathrm{H}$ in the gas-phase. Previous observational studies that have detected \SO and \SOtwo (and their isotopologues, but no \HtwoS) in warm Herbig disks thus suggested a potential link to the sublimation of \HtwoS and H$_2$O ices \citep{Booth2021, Booth2023, Booth2023_HD169142, Booth2025}. In the present observations of the HD~163296 disk, the emitting radius of \SO (and presumably \SOtwo) are similar to that of \HtwoS, supporting gas-phase formation of \SO and \SOtwo via the reactions above. The lack of the CS emission in the SO-emitting radius (see Section \ref{subsubsec:sulfur_inner_outer}) may also support the gas-phase formation of SO and \SOtwo, possibly in the disk surface, where CS is destroyed by strong UV radiation and further facilitate the formation of SO and \SOtwo using the atomic sulfur originated from CS. The slightly larger radius of \SO (and \SOtwo) than \HtwoS could be explained by the gas excitation effect, i.e., the lower upper state energy of the \SO line ($E_\mathrm{u}\approx67$\,K) than that of the \HtwoS line ($E_\mathrm{u}\approx169$\,K), or additional contributions from \SO and \SOtwo ices that have lower binding energies than that of water and thus desorb at a lower temperature ($\lesssim100$\,K). Indeed, experimental studies suggest that the trapping of \SOtwo within water ice is less significant compared to \HtwoS case \citep{Collings2004}. All together, the observed emission of sulfur-bearing molecules in the HD~163296 disk can be explained by the sublimation of sulfur-rich ices and potential subsequent gas-phase processing in the inner region of the disk.

% slightly larger than that of \HtwoS. Although this could be due to gas excitation effect, i.e., due to the lower upper state energy of the \SO line ($E_\mathrm{u}\approx67$\,K) than that of the \HtwoS line  ($E_\mathrm{u}\approx169$\,K), \SO and \SOtwo ices may desorb from grain surface at lower temperature ($\lesssim100$\,K) as they have lower binding energies than that of water, and the warm gas of \SO and \SOtwo can thus distribute at larger radii. Indeed, experimental studies suggest that the trapping of \SOtwo within water ice is less significant compared to \HtwoS case \citep{Collings2004}. All together, the observed emission of sulfur-bearing molecules in the HD~163296 disk can be explained by the sublimation of sulfur-rich ices and potential subsequent gas-phase processing in the inner region of the disk.
% , providing first direct evidence of ice sublimation by \HtwoS detection.   

\subsubsection{Possible Link to the Jet/Outflow Activities}
The HD~163296 system is known to host optical jets and molecular outflows in the innermost region ($\lesssim10$\,au) of the disk. The collimated optical jets, named as HH~409, show high-velocities ($\approx200$\,km\,s$^{-1}$) and knotty structures \citep{Wassell2006, Ellerbroek2014}, while the molecular outflows have been detected in CO molecular line emission at lower velocities ($\approx20$\,km\,s$^{-1}$; \citealt{Klaassen2013, Booth2021_HD163296}). The molecular outflows are thought to be a magnetohydrodynamical (MHD) disk wind with an estimated launching radius of $\approx4$\,au \citep{Booth2021_HD163296}, which is remarkably similar to the emitting radii derived in the present analysis. While it is unlikely that the line emission in the present work traces the disk wind itself as the velocity structures are different from that of CO in \citet{Booth2021_HD163296}, this coincidence may imply that it traces the launching region of the disk wind at the disk surface, where disk materials can be heated. Particularly, the gas-phase products (\SO and \SOtwo) may trace near the disk surface where the destruction of molecules by UV radiation is more efficient than midplane (see Section \ref{subsubsec:S_ice_sublimation}), but additional observations and/or more sophisticated modeling are needed to understand their exact physical connections to the disk wind and/or the jet activity.

% It is also noteworthy that the emitting radii of these molecules are coincide well with the launching radius ($\sim4$\,au; \citealt{Booth2021_HD163296}) of the known molecular outflows in the HD~163296 system \citep{Wassell2006, Ellerbroek2014}. The molecular outflow is thought to be an magnetohydrodynamical (MHD) disk wind \citep{Booth2021_HD163296}. This coincidence may thus imply that \HtwoS, \SO, and \SOtwo trace the launching region of the MHD disk wind at the disk surface. P 

\subsection{Sulfur Reservoir in the HD 163296 Disk}

\subsubsection{Gas-phase Sulfur Partitioning in the Inner and Outer Disk}\label{subsubsec:sulfur_inner_outer}
In the present analysis, column density and temperature are highly degenerate for all species as shown in Figure \ref{fig:N-T_plot}. In Figure \ref{fig:N-T_plot}, we indicate the midplane temperature at the best-fit emitting radius for each molecule expected for a passively heated, flared disk in radiative equilibrium \citep[e.g.,][]{Chiang1997} as a reference;
\begin{equation}
    T_\mathrm{mid}(r) = \left(\frac{\varphi L_\star}{8 \pi r^4 \sigma_\mathrm{SB}}\right)^{0.25},
\end{equation}
where $\varphi$ is the flaring angle, $L_\star$ is the stellar luminosity, and $\sigma_\mathrm{SB}$ is the Stefan-Boltzmann constant. We use the known HD~163296's bolometric luminosity of $17\,L_\odot$ as $L_\star$ and assume $\varphi = 0.02$ as in \citet{Huang2018}. At the best-fit emitting radii of $\approx3$\,au and $\approx5$\,au, the midplane temperatures are $\approx150$\,K and $\approx120$\,K, respectively. With these temperatures the solutions are in optically thick regime (see Figure \ref{fig:N-T_plot}) and thus only lower limits of column densities can be inferred. The sulfur partitioning between different molecules are therefore uncertain at this regime. On the other hand, column densities can be better constrained for temperatures of $\gtrsim300$\,K, which correspond to the optically thin limit. Such a high temperature would correspond to the disk surface \citep[$z/r\sim0.2$ at a radius of $\sim3$--5\,au;][]{Pirovano2022}, and it may be possible that the observed emission originates from there.
% While only the lower limit of column densities can be inferred at low temperature ($\lesssim300$\,K) and the sulfur partitioning between these molecules are thus uncertain at this regime, column densities can be better constrained for temperatures of $\gtrsim300$\,K, which corresponds to the optically thin limit. Such a high temperature would correspond to the disk surface \citep[$z/r\sim0.2$ at $\sim3$--5\,au radius;][]{Pirovano2022}, and it may be possible that the observed emission originates from there. 
% if dust grains have flared distributions, given the icy origin of the detected molecules. 
For this regime, we tentatively derived the column density ratios of \SO and \SOtwo relative to \HtwoS for the posterior MCMC samples with $T>300$\,K to be 0.13 and 0.46, respectively, indicating that \HtwoS could be the potential major gas-phase sulfur carrier in the inner disk. This is broadly consistent with disk chemical models that predict \HtwoS to be one of the most abundant sulfur-bearing volatile species \citep[e.g.,][]{Dutrey2011, Semenov2018, Riviere-Marichalar2022}, and with the fact that \HtwoS is the most abundant sulfur-bearing molecules in cometary ice \citep[e.g.,][]{Bockelee-Morvan2000, Calmonte2016}.

\begin{figure}
% \epsscale{1.15}
\plotone{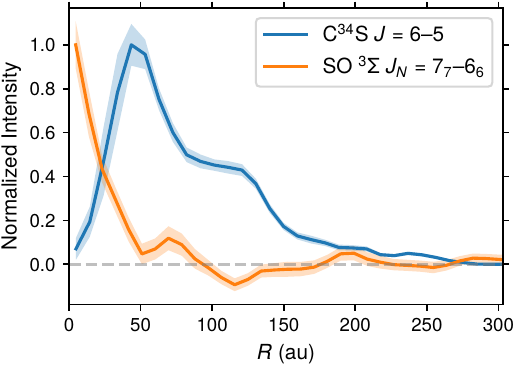}
\caption{Deprojected radial intensity profiles of C$^{34}$S $J=6$--5 (blue) and SO $^3\Sigma\,J_N=7_7$--$6_6$ (orange). The C$^{34}$S line is observed in the same project as the SO line, and its radial profile is taken from \citet{Law2025}. These profiles are generated by the the \texttt{radial\_profile} function built in \texttt{GoFish} \citep{GoFish}, assuming a flat (i.e., $z/r = 0$) emission surface. The profiles are normalized by their peak value. The shaded region indicates $1\sigma$ uncertainty.   
}
\label{fig:radial_profile}
\end{figure}

% recent chemical modeling effort tailored to the disk around another famous  Herbig Ae star HD~100546 have found that \HtwoS is not a major sulfur reservoir in both gas- and solid-phases throughout the disk \citep{Keyte2024}. Their model instead predicts OCS to be an abundant grain-surface species.
% , although an experimentally-proposed pathway that efficiently convert OCS to \HtwoS on grain surface \citep{Nguyen2021} is not included in their chemical network

In the outer disk of HD~163296, sulfur-bearing molecules that contain carbon are strongly detected \citep[][see also Appendix \ref{appendix:other_lines} for a detection of rare $^{34}$S isotopolgues of H$_2$CS]{LeGal2019, Law2025}. \citet{Law2025} compiled multi-line archival data of CS, H$_2$CS, and their $^{34}$S isotopologues, and found they take the form of multi-ringed distributions in the outer ($\gtrsim30$\,au) disk. While these molecules are abundant and thus serves as major gas-phase sulfur reservoirs in the outer region, \textrm{the radial intensity profiles of CS and its isotopologue lines (with $E_\mathrm{u}\approx7$--85\,K)} consistently show decreasing trend toward the disk center at $\lesssim30$\,au, where SO emission is present. Figure \ref{fig:radial_profile} shows the deprojected radial intensity profiles of the C$^{34}$S $J=6$--5 line (taken from \citealt{Law2025}) and the \SO line in the HD~163296 disk, showcasing a clear anti-correlation in radial distributions. \textrm{Although the radial intensity profiles may be affected by disk physical conditions,} this likely points to a change in the CS/SO abundance ratio, and in turn, C/O ratio \citep{Semenov2018, LeGal2021, Keyte2024, Williams2025}. Similar radial segregation between carbonated and oxygenated sulfur-bearing species have also been observed in other Herbig disks as well \citep[e.g.,][]{Booth2023_HD169142, Booth2024_HD100546}.
% , and likely point to a change in the gas-phase C/O ratio. 
% due to the sublimation of water ice, one of the major oxygen carrier. 
We note that, in the HD~163296 disk, there is a gap between the radii where the SO emission is present ($\sim5$\,au) and the CS and H$_2$CS emission disappears ($\sim30$\,au), but this can be explained by the optically thick dust continuum emission at $\lesssim30$--40\,au as suggested by a multi-wavelength study \citep{Guidi2022}, which obscure the line emission there. The reason why we detect warm gas emission in these inner region could be due, for example, to elevated molecular emitting layers, and/or to an unresolved inner dust cavity at $\lesssim5$\,au that reduces the dust optical depth. Indeed, radial intensity profiles of (sub-)mm continuum emission derived from visibility analysis by \citet{Guidi2022} show a hint of central depression at $\lesssim5$\,au. In any case, there is likely radial and/or vertical variations in the CS/SO ratio (and thus C/O ratio) in the innermost region of the disk ($\lesssim30$\,au).

% Figure \ref{} shows the radial intensity profiles of SO, C$^{34}$S $J=6$--5, and H$_2$C$^{34}$S $J_{K_a, K_c} = 9_{1,9}$--$8_{1,8}$, all of which are simultaneously observed in the same dataset (see also \citealt{Law2025})  

\subsubsection{Abundances Relative to Water}\label{subsubsec:abundance_to_water}
\begin{figure}
\epsscale{1.1}
\plotone{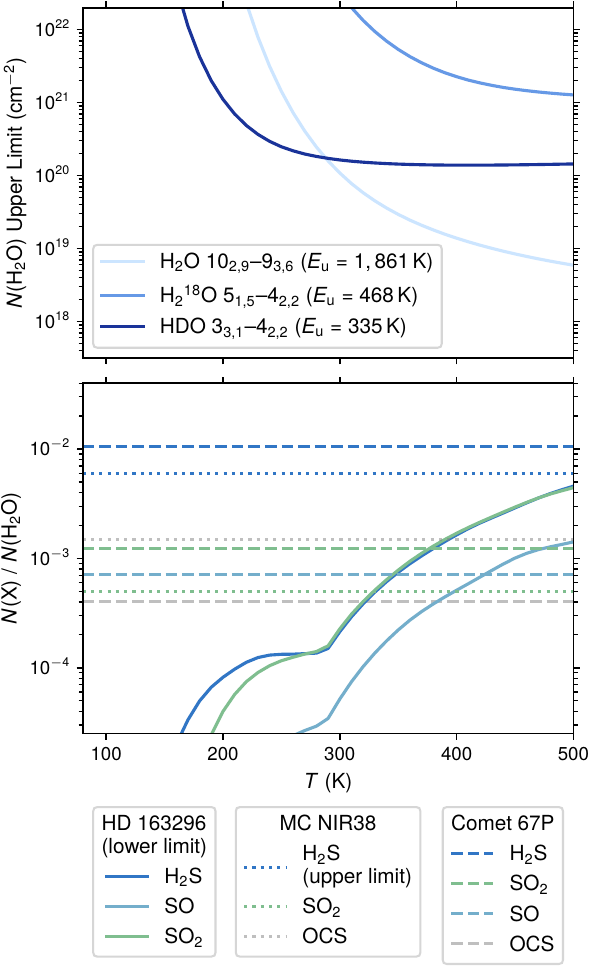}
\caption{Upper limits on water column density (top) and lower limits on the abundance ratios of detected sulfur-bearing species relative to water (bottom, solid lines) as a function of assumed gas temperature. The different colors on the top panel indicate the constraints from different isotopologue lines. The lower limits on the abundance ratios correspond to the lower end of the $1\sigma$ confidence intervals in Figure \ref{fig:N-T_plot}. 
% The filled areas in the bottom panel correspond to the $1\sigma$ confidence intervals in Figure \ref{fig:N-T_plot}. 
% The ratios are not well constrained in the low temperature regime ($\lesssim160$\,K) due to the divergence of column densities. 
The colored horizontal lines on the bottom panel marks the ice abundance ratios of \HtwoS, \SOtwo, \SO, \textrm{and OCS} in the molecular cloud NIR38 (dotted; \citealt{McClure2023}) and in comet 67P/C-G (dashed; \citealt{Calmonte2016}) as indicated by the legend. 
% The gray horizontal lines indicate the abundance ratios of OCS in molecular cloud (dotted) and comet (dashed) as a reference. 
Note that in the molecular cloud, the \HtwoS abundance ratio is an upper limit, and no constraints on SO ice have been reported.
}
\label{fig:abundance_comparison}
\end{figure}

Given that the molecules detected in this study could have an ice origin, it would be informative to calculate the abundance relative to water and compare it with the ice abundance in different evolutionary stages. \citet{Notsu2019} observed ortho-H$_2$O, para-H$_2$$^{18}$O, and HDO lines in ALMA Band 7 toward the HD~163296 disk, but none of them were detected. We reanalyzed this archival data\footnote{We used the Additional Representative Images for Legacy (ARI-L) products available at the ALMA Science Archive.} (project code: 2015.1.01259.S) to derive the $3\sigma$ upper limits of integrated flux densities, assuming a line width of 40 km\,s$^{-1}$, to be 69, 96, and 48 \,mJy\,km\,s$^{-1}$ for H$_2$O, H$_2$$^{18}$O, and HDO lines, respectively\footnote{Note that, unlike \citet{Notsu2019} where they assume a 20\,au emitting radius, here we assume a point source emission given that the postulated emitting radius (same as that of \HtwoS; $\sim3$\,au or 0\farcs03) is smaller than the beam size ($\sim0\farcs1$).}. Using these upper limits, we computed the upper limits of the water column densities for a range of temperature (Figure \ref{fig:abundance_comparison}), assuming the same emitting radius as that of \HtwoS ($\sim$3\,au; see Table \ref{tab:lineprofilefits}), a thermalized ortho-to-para ratio of three for H$_2$O and H$_2$$^{18}$O, and the standard isotopologue ratios of H$_2$$^{16}$O/H$_2$$^{18}$O $=557$ \citep{Wilson1999} and HDO/H$_2$O $=2\times10^{-3}$ \citep{Tobin2023, Andreu2023}. The upper limits are sensitive to the assumed temperature, where they are effectively constrained from the HDO and H$_2$O lines at lower and higher temperature regimes, respectively. 

Combining with the column constraints on \HtwoS, \SOtwo, and \SO, we also compute the lower limits of the abundance ratios relative to water, and compare them with those in the molecular cloud NIR38 \citep{McClure2023} and comet 67P/C-G \citep{Calmonte2016} in Figure \ref{fig:abundance_comparison} (bottom). While \HtwoS abundance is broadly consistent with both the cometary value and the molecular cloud upper limit, \SO and \SOtwo abundances may be higher than those in molecular cloud and comet when a high temperature ($\gtrsim400$\,K) is assumed. This might indicate that \SO and \SOtwo abundances are enhanced by grain-surface and/or warm gas-phase processing in the protoplanetary disk stage \citep[e.g.,][]{Semenov2018, Keyte2024}, but better constraints on temperature and column density with multi-line observations are needed for a firm conclusion. Moreover, the current constraints suggest that the total volatile sulfur abundance in the HD~163296 disk is $\gtrsim10^{-8}$--$10^{-6}$ (assuming an H$_2$O abundance of 10$^{-4}$), which do not contradict with those in the ISM and comets (Figure \ref{fig:abundance_comparison}). 
% Considering that the H$_2$S abundance here may include the sublimation product of NH$_4$$^+$SH$^-$, 
This is broadly compatible with the scenario that the volatile sulfur is depleted in disks into more refractory materials, such as sulfide minerals (e.g., FeS) and salts (e.g., NH$_4$$^+$SH$^{-}$), as suggested by previous studies \citep{Keller2002, Kama2019, Keyte2024}.

Another potentially important molecule for tracing the chemical evolution of sulfur is OCS, which is one of the sulfur-bearing ices detected in molecular clouds and young stellar objects \citep[e.g.,][]{Palumdo1995, Palumdo1997, Aikawa2012, McClure2023}.
% but not in protoplanetary disks \citep[][see Section \ref{subsec:herbig_disks}]{Booth2024}. 
As shown in the bottom panel of Figure \ref{fig:abundance_comparison}, OCS ice is present at $0.05$--$0.1\%$ level of water in molecular clouds and comets \citep{McClure2023, Calmonte2016}, which is comparable to \SOtwo and \SO. \citet{Drozdovskaya2018} found that the warm inner envelope of the Class 0 protostar IRAS 16293-2422 B, where ices have sublimated, contains much more OCS than \HtwoS compared to comet 67P/C-G. They interpreted this as an outcome of the different levels of UV radiation during core collapse, where a higher level of UV radiation will more efficiently convert \HtwoS to OCS. A few recent observational studies also suggest that the \HtwoS/OCS ratio can trace the different birth environments of protostars, including the level of UV radiation and heating \citep{Kushwahaa2023, Miranzo-Pastor2025}. \textrm{As for more evolved protoplanetary disks, OCS is abundant with an \SOtwo/OCS ratio of $\sim$0.6 in the V883 Ori disk \citep{Yamato2024_V883Ori}, which likely trace the ice composition, whereas the amount of OCS is less than a few \% of \SOtwo in transition disks \citep[][see Section \ref{subsec:herbig_disks}]{Booth2024_IRS48, Booth2024_HD100546}.} Observations of OCS, as well as multi-line observations that will help better constrain the abundance of sulfur-bearing molecules, will be key in better understanding the sulfur evolution during protoplanetary disk stage.

\subsection{In the Context of Other Herbig Disks}\label{subsec:herbig_disks}
Recent sensitive ALMA observations have revealed warm sulfur-bearing molecular gas in a handful of disks around Herbig Ae/Be stars \citep[e.g.,][]{Booth2021, Booth2023, Booth2023_HD169142, Booth2024_IRS48, Booth2024_HD100546, Booth2026}. These disks are all transition disks with a sizable central cavity \textrm{($\sim10$--20\,au)}, and are associated with warm \textrm{($\gtrsim100$\,K)} COM emission, which is likely an outcome of ice sublimation at the irradiated cavity edge. In these disks, particularly well-studied IRS 48 and HD 100546 disks, \SO and \SOtwo are strongly detected and shows high column densities ($\sim10^{14}$--10$^{15}$\,cm$^{-2}$), but no OCS emission is detected and its column density is less than $\sim1$\% of SO \citep{Booth2024_HD100546, Booth2024_IRS48}. \textrm{These column densities of SO and \SOtwo are lower than those in HD~163296 shown here ($>10^{15}$\,cm$^{-2}$), but this is due to the difference in the amount of total gas rather than chemistry, as they trace different disk radii (up to $\sim100$\,au and $\sim3$--5\,au).} As for \HtwoS, while no \HtwoS lines are observed in the recent, sensitive ALMA spectral survey data of HD~100546 and IRS~48 \citep{Booth2024_HD100546, Booth2024_IRS48}, an old Cycle 0 dataset toward HD~100546 disk (project code: 2011.0.00863.S), which detected SO emission \citep{Booth2018}, covers the same \HtwoS transition observed in the present study, but shows no detection\footnote{We confirmed this by investigating the data products available at the ALMA Science Archive.}. A similar trend has been seen in a recent transition disk survey by \citet{Booth2025}, where they found a weak H$_2$S detection only in one disk out of six observed disks. \textrm{In these disks, the line fluxes of SO and \SOtwo are typically higher than the upper limits of \HtwoS fluxes \citep{Booth2025}, implying that the amount of SO and \SOtwo could be higher than that of \HtwoS.} This is in contrast to the \textit{full} disk around HD~163296 where both SO and \HtwoS has been detected at a similar brightness \textrm{and \HtwoS could have a higher column density (see Section \ref{subsubsec:sulfur_inner_outer})}. Although this difference may not be robust due to low S/Ns, sulfur reservoir in the HD~163296 disk may experience less substantial reprocessing compared to the transition disks, where UV radiation from the central star can directly irradiate the ices at cavity edge and destroy and convert \HtwoS, which likely only have a low-temperature grain-surface formation pathway, to other molecules, such as \SO and \SOtwo \citep[e.g.,][]{Semenov2018, Keyte2024}. \textrm{High \SO/\SOtwo ratios of $1$--2, which indicate photodissociation of \SOtwo that forms SO, are also reported in the transitions disks \citep{Booth2021, Booth2024_IRS48, Booth2024_HD100546}, while HD~163296 shows a (tentative) lower \SO/\SOtwo ratio of $\sim0.3$ (see Section \ref{subsubsec:sulfur_inner_outer}), consistent with the scenario above.} 
% Alternatively, as suggested in the previous study \citep{Booth2023}, it might also be possible that the local shock around the protoplanet embedded in the HD~100546 disk selectively enhance the abundance of \SO and \SOtwo, which are well-known shock tracers. 
\textrm{Given that all the abundance ratio measurements in the HD~163296 disk are tentative, and} there are already some other hints of varying degree of chemical reprocessing between \textit{full} disks and \textit{transition} disks \citep{Yamato2024}, further observations toward these different types of disks will shed light on the effect of the radiation from the central star on the evolution of sulfur content in disks.

\section{Summary} \label{sec:summary}

We have presented sensitive ALMA observations of sulfur-bearing molecules toward the HD~163296 disk, which have detected compact emission of \HtwoS, \SO and tentatively \SOtwo in the central region of the disk. We summarize our main findings as follows: 
\begin{enumerate}
    \item We detected a compact, unresolved emission of a \HtwoS line with a high upper state energy ($E_\mathrm{u}\approx169$\,K) at $\sim0\farcs3$ (or $\sim30$\,au) resolution, as well as \SO and tentative \SOtwo emission in the central region of the HD~163296 disk. This is similar to the emission of the same \HtwoS line recently detected in the HD~169142 disk \citep{Booth2025}, but in contrast to the previous studies showing ring-like emission distributions of a low-excitation line in the outer disk.
    \item The \HtwoS and \SO emission show broad line widths of $\sim40$\,km\,s$^{-1}$, indicating an inner disk origin. We constrained the emitting radius, gas temperature, and column density of \HtwoS, \SO, and \SOtwo via line profile fits with a Keplerian-rotating disk model, where the emitting radii are derived to be $\sim3$--5\,au. While the gas temperature and column densities are degenerate, their lower limits are roughly constrained to be $\gtrsim100$\,K and $\gtrsim10^{16}$\,cm$^{-2}$, indicating the emission traces the inner warm region.
    \item We suggest that the majority of the observed emission arises from the sulfur-bearing ice sublimation, associated with water ice sublimation, based on high gas temperature and comparable emitting radii with the water snowline radius as constrained in previous studies. While \SO and \SOtwo may indicate some degree of gas-phase processing as well, \HtwoS provides evidence of ice origin, \textrm{as previous astrochemical models suggest that it only forms on dust grain surfaces.}
    \item We tentatively derive the \SO/\HtwoS and \SOtwo/\HtwoS ratios to be 0.13 and 0.46, suggesting that \HtwoS could be the most abundant gas-phase volatile sulfur species in the inner region of the HD~163296 disk. This needs to be confirmed with higher-sensitivity, multi-line observations that will help better constrain the abundance ratios. 
    % We also show that the inner disk of HD~163296 is likely O-rich environment unlike the high-C/O outer disk.
    \item Utilizing ALMA archival data, we also compute the abundance of sulfur-bearing molecules relative to water and compare them with the ISM and comets. Within the current, limited constraints, the abundance ratios are broadly consistent with both ISM and comets. This is broadly compatible with the literature predictions that the volatile sulfur is depleted into more refractory species in protoplanetary disks as well.
    \item We compare the HD~163296 disk with transition disks that show signatures of ice sublimation. The similar brightness of \HtwoS to that of SO in the HD~163296 \textit{full} disk, which is in contrast to the recent transition disk observations, may reflect the different degree of in-situ processing by UV radiation caused by the absence/presence of the central dust cavity. 
\end{enumerate}

% \begin{acknowledgments}
\textrm{The authors thank the anonymous referee for their comments that improved the content of this work.} This paper makes use of the following ALMA data: ADS/JAO.ALMA\#2021.1.00535.S. ALMA is a partnership of ESO (representing its member states), NSF (USA) and NINS (Japan), together with NRC (Canada), NSTC and ASIAA (Taiwan), and KASI (Republic of Korea), in cooperation with the Republic of Chile. The Joint ALMA Observatory is operated by ESO, AUI/NRAO and NAOJ.
Y.Y. is financially supported by Grant-in-Aid for the Japan Society for the Promotion of Science (JSPS) Fellows (KAKENHI Grant Number JP23KJ0636 and JP25K23409) and the RIKEN Special Postdoctoral
Researcher Program (Fellowships). Support for C.J.L. was provided by NASA through the NASA Hubble Fellowship grant Nos. HST-HF2-51535.001-A awarded by the Space Telescope Science Institute, which is operated by the Association of Universities for Research in Astronomy, Inc., for NASA, under contract NAS5-26555.
R.L.G. acknowledges support from the French Agence Nationale de la Recherche (ANR) through the project MAPSAJE (ANR-24-CE31-2126-01).
S.N. is grateful for support from Grants-in-Aid for JSPS (Japan Society for the Promotion of Science) Fellows grant No. JP23KJ0329, MEXT/JSPS Grants-in-Aid for Scientific Research (KAKENHI) grant Nos. JP23K13155, JP24K00674, and JP23H05441, and Start-up Research Grant as one of The University of Tokyo Excellent Young Researcher 2024. V.V.G. acknowledge support from the ANID -- Millennium Science Initiative Program -- Center Code NCN2024\_001, from FONDECYT Regular 1221352, and ANID CATA-BASAL project FB210003.
% \end{acknowledgments}


\vspace{5mm}
\facilities{ALMA}


\software{CASA \citep{CASA}, \texttt{bettermoments} \citep{bettermoments}, \texttt{emcee} \citep{emcee}, \texttt{GoFish} \citep{GoFish}}
 


\appendix

\section{Additional Line Detections}\label{appendix:other_lines}
\begin{figure*}
% \epsscale{1.15}
\plotone{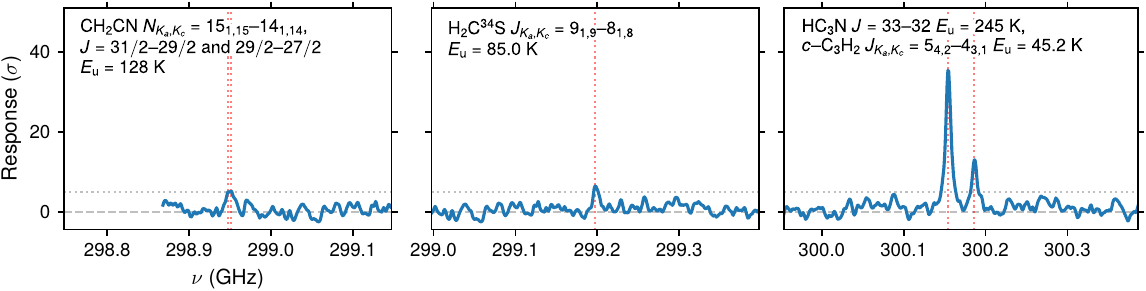}
\caption{Matched filter responses for molecular lines with $>5\sigma$ signal in the continuum spectral window. The horizontal gray dotted line marks $5\sigma$ level. The expected line frequencies are indicated by the vertical red dotted lines.
}
\label{fig:matched_filter_other}
\end{figure*}

\begin{figure}
\epsscale{1.15}
\plotone{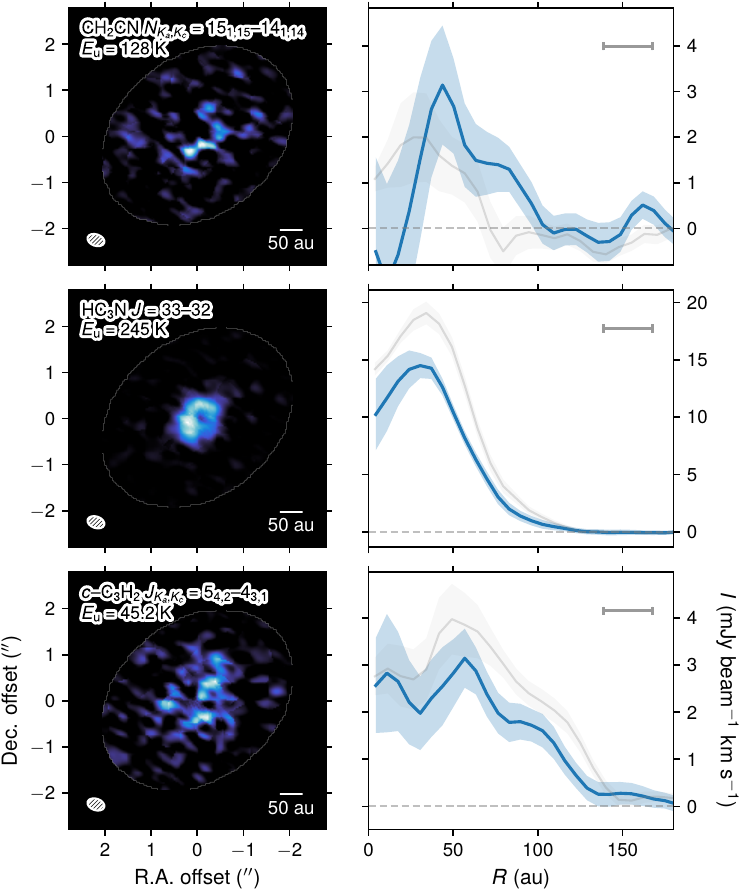}
\caption{Velocity-integrated intensity maps (left) and deprojected radial intensity profiles (right) of CH$_2$CN (top), HC$_3$N (middle), and $c$-C$_3$H$_2$ (bottom). In the left panels, synthesized beams and 50 au scale bars are shown in the bottom left and right corners, respectively. In the right panels, blue-shaded regions indicate 1$\sigma$ uncertainties. The size of the beam major axis is indicated in the top right corner. As a reference, radial intensity profiles of the chemically-related species or the same molecule's transitions with a similar excitation condition (i.e., similar upper state energy) are shown in the light gray curves; CH$_3$CN $J_K = 12_3$--$11_3$ ($E_\mathrm{u} = 133$\,K) for CH$_2$CN (top), HC$_3$N $J=29$--28 ($E_\mathrm{u}=190$\,K) for HC$_3$N (middle), and $c$-C$_3$H$_2$ $J_{K_a,K_c}=6_{1,5}$--$5_{2,4}$ ($E_\mathrm{u} = 47.5$\,K) for $c$-C$_3$H$_2$ (bottom). These profiles are taken from the website of the Molecules with ALMA at Planet-forming Scales (MAPS) ALMA Large Program (\url{https://alma-maps.info/}).
}
\label{fig:ancillary_lines}
\end{figure}

In addition to the \HtwoS, \SO, and \SOtwo lines, which are the main focus of this study, the same dataset includes detections of a few additional molecules thanks to the deep nature of the observations. Here we present molecular line detections in the wide continuum spectral window and their basic analysis. To blindly search for line detections, we performed a matched filter analysis using a filter based on a simple Keplerian-rotating velocity pattern with known disk geometries and an inner and outer radius of 30 and 70\,au, respectively, which covers the region where the majority of molecular line emission arises from. Figure \ref{fig:matched_filter_other} shows the filter responses for molecular lines with $>5\sigma$ signal. We detect CH$_2$CN, H$_2$C$^{34}$S, HC$_3$N and $c$-C$_3$H$_2$, among which H$_2$C$^{34}$S is the first-ever detection in a protoplanetary disk (see also \citealt{Law2025}), and CH$_2$CN is the second disk detection following the detection in the TW~Hya disk \citep{Canta2021}. The CH$_2$CN signal is a blend of two individual hyperfine lines. Table \ref{tab:ancillary_lines} lists the detected molecular lines, their spectroscopic properties, and flux densities calculated by spatially integrating the velocity-integrated intensity maps (Figure \ref{fig:ancillary_lines}).

Figure \ref{fig:ancillary_lines} shows the velocity-integrated intensity maps and deprojected radial intensity profiles for the CH$_2$CN, HC$_3$N and $c$-C$_3$H$_2$ lines (see \citealt{Law2025} for the same analysis of the H$_2$C$^{34}$S line). These maps and radial profiles are generated using the Python package \texttt{bettermoments} \citep{bettermoments} with a Keplerian mask and the \texttt{radial\_profile} function built in the Python package \texttt{GoFish} \citep{GoFish}. The radial intensity profiles of HC$_3$N and $c$-C$_3$H$_2$ show similar morphologies to those of the same molecule's different transitions, which are originally published by \citet{Ilee2021}. As for CH$_2$CN, which is initially detected in this disk in the present study, the radial distributions could be slightly extended compared to that of the chemically related species, CH$_3$CN \citep{Ilee2021}, although S/N is low. Similar radial distributions and excitation temperatures have been found for CH$_2$CN and CH$_3$CN in the TW~Hya disk \citep{Canta2021}. 

We further roughly quantify the disk-averaged column density of CH$_2$CN assuming (1) the same emitting region as that of CH$_3$CN \citep{Ilee2021}, (2) the same disk-averaged excitation temperature (35\,K) as that of CH$_3$CN in LTE condition \citep{Ilee2021}, and (3) fully optically thin CH$_2$CN emission. Using the standard equation for the optically thin emission in LTE \citep[see e.g.,][]{Canta2021} and taking the hyperfine splitting of the CH$_2$CN line into account, we derived a disk-averaged CH$_2$CN column density of $\approx1.1\times10^{13}$\,cm$^{-2}$. This is higher than the disk-averaged CH$_3$CN column density reported in \citet{Ilee2021} ($\approx2.3\times10^{12}$\,cm$^{-2}$) by a factor of $\approx5$, i.e., CH$_2$CN/CH$_3$CN $\approx5$. This high ratio is consistent with the ratio found in the TW~Hya disk ($\approx4$--10; \citealt{Canta2021}), suggesting that high CH$_2$CN/CH$_3$CN ratios may be common in protoplanetary disks.

\begin{deluxetable*}{lccccc}
\label{tab:ancillary_lines}
% \tablewidth{\textwidth}
\tablecaption{Detected Additional Molecular Lines}
\tablehead{\colhead{Transition} & \colhead{$\nu_0$} & \colhead{$E_\mathrm{u}$} & \colhead{$\log_{10}A_\mathrm{ul}$}  & \colhead{$g_\mathrm{u}$} & \colhead{$F_\nu$} \\
\colhead{} & \colhead{(GHz)} & \colhead{(K)} & \colhead{(s$^{-1}$)} & \colhead{} & \colhead{(mJy\,km\,s$^{-1}$)}} 
\startdata
CH$_2$CN $N_{K_a, K_c} = 15_{1,15}$--$14_{1,14}, J=31/2$--$29/2$ & 298.952689 & 128.0 & $-$2.736 & 96 & \multirow{2}{*}{$27^\dagger$} \\
CH$_2$CN $N_{K_a, K_c} = 15_{1,15}$--$14_{1,14}, J=29/2$--$27/2$ & 298.957008 & 128.0 & $-$2.737 & 90 &  \\
H$_2$C$^{34}$S $J_{K_a, K_c} = 9_{1,9}$--$8_{1,8}$ & 299.203225 & 85.0 & $-$3.402 & 57 & $52^\ddagger$ \\
HC$_3$N $J = 33$--$32$ & 300.159647 & 244.9 & $-$2.666 & 67 & $95$ \\
$c$-C$_3$H$_2$ $J_{K_a, K_c} = 5_{4,2}$--$4_{3,1}$ & 300.191718 & 45.3 & $-$3.160  & 11 & $40$
\enddata
\tablenotetext{}{$^\dagger$Sum of the flux densities for the two hyperfine CH$_2$CN transitions.}
% \tablecaption{$^\dagger$}
\tablenotetext{}{$^\ddagger$Taken from \citet{Law2025}.}
\tablecomments{The spectroscopic properties of the lines are taken from the Cologne Database for Molecular Spectroscopy (CDMS; \citealt{CDMS1, CDMS2, CDMS3}) with the original data from \citet{Saito1997} for CH$_2$CN, from \citet{Mullar2019} for H$_2$C$^{34}$S, from \citet{Thorwirth2000} for HC$_3$N, and \citet{Bogey1986} for $c$-C$_3$H$_2$.}
\end{deluxetable*}

\section{\SOtwo Line Stacking}\label{appendix:SO2_stacking}
\begin{figure}
\epsscale{0.95}
\plotone{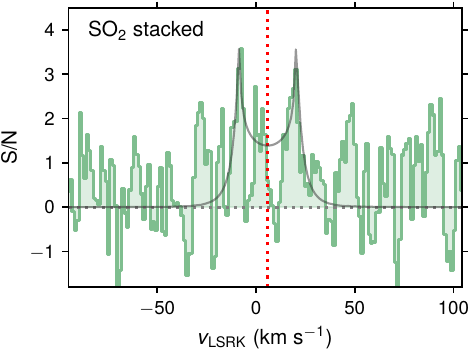}
\caption{\SOtwo spectrum stacked over two observed transitions. The vertical red dotted line marks the systemic velocity ($\approx5.76$\,km\,s$^{-1}$), while the horizontal gray dotted line indicate the zero level. The best-fit Keplerian-rotating disk model (Section \ref{sec:analysis}) are shown in the gray solid curve as a guide.}
\label{fig:SO2_stacked}
\end{figure}

Figure \ref{fig:SO2_stacked} shows the \SOtwo spectrum stacked over two observed transitions as shown in Figure \ref{fig:spectra}. We employed the standard S/N-weighted stacking. The stacked spectrum are normalized by the RMS noise measured in the line-free region, i.e., $<-50$\,km\,s$^{-1}$ and $>50$\,km\,s$^{-1}$. The stacked spectrum shows a $\approx3.6\sigma$ signal and roughly follows the best-fit Keplerian-rotating disk model (see Section \ref{sec:analysis}) as guided by the gray solid line.

\section{Corner plot of the MCMC fit}\label{appendix:corner_plot}
\begin{figure*}
\epsscale{1.15}
\plotone{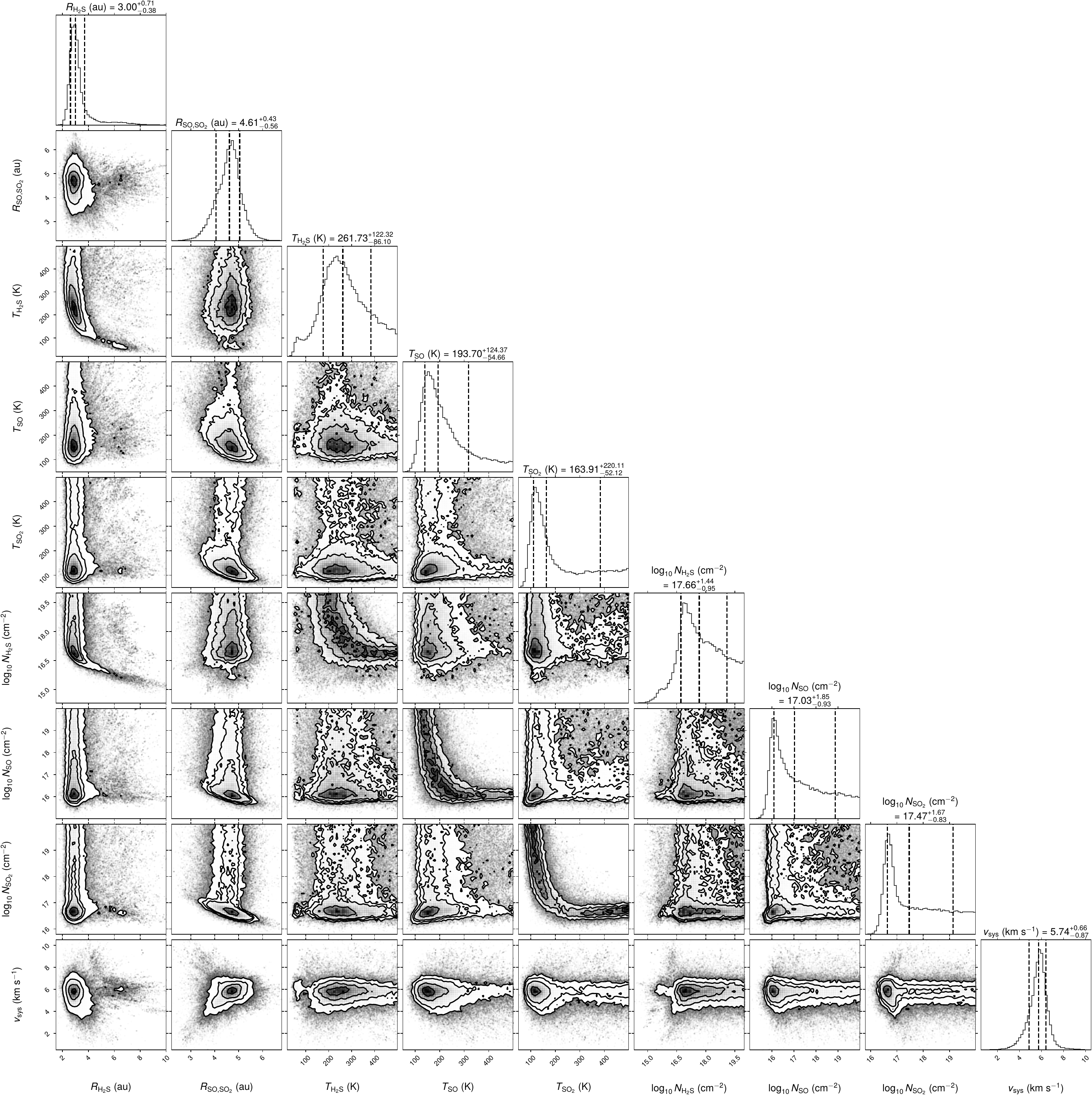}
\caption{Marginalized posterior distributions and covariances between different parameters of the MCMC fit to the line profiles. The vertical dashed lines in each panel that show the marginalized posteriors distribution marks 16th, 50th, and 84th percentiles. The covariances between column density ($\log_{10}N$) and temperature ($T$) for each molecular species are also shown in Figure \ref{fig:N-T_plot}.
}
\label{fig:corner_plot}
\end{figure*}

Figure \ref{fig:corner_plot} shows the marginalized posterior distributions and covariances between different parameters of the MCMC fit to the line profiles.  

% \section{Other Line Detections}


\bibliography{sample631}{}

@ARTICLE{Guidi2022,
       author = {{Guidi}, G. and {Isella}, A. and {Testi}, L. and {Chandler}, C.~J. and {Liu}, H.~B. and {Schmid}, H.~M. and {Rosotti}, G. and {Meng}, C. and {Jennings}, J. and {Williams}, J.~P. and {Carpenter}, J.~M. and {de Gregorio-Monsalvo}, I. and {Li}, H. and {Liu}, S.~F. and {Ortolani}, S. and {Quanz}, S.~P. and {Ricci}, L. and {Tazzari}, M.},
        title = "{Distribution of solids in the rings of the HD 163296 disk: a multiwavelength study}",
      journal = {\aap},
         year = 2022,
        month = aug,
       volume = {664},
          eid = {A137},
        pages = {A137},
          doi = {10.1051/0004-6361/202142303},
archivePrefix = {arXiv},
       eprint = {2207.01496},
 primaryClass = {astro-ph.EP},
       adsurl = {https://ui.adsabs.harvard.edu/abs/2022A&A...664A.137G}
}

@ARTICLE{Pirovano2022,
       author = {{Pirovano}, L.~M. and {Fedele}, D. and {van Dishoeck}, E.~F. and {Hogerheijde}, M.~R. and {Lodato}, G. and {Bruderer}, S.},
        title = "{H$_{2}$O distribution in the disc of HD 100546 and HD 163296: the role of dust dynamics and planet-disc interaction}",
      journal = {\aap},
         year = 2022,
        month = sep,
       volume = {665},
          eid = {A45},
        pages = {A45},
          doi = {10.1051/0004-6361/202244104},
archivePrefix = {arXiv},
       eprint = {2207.10744},
 primaryClass = {astro-ph.EP},
       adsurl = {https://ui.adsabs.harvard.edu/abs/2022A&A...665A..45P}
}

@ARTICLE{bettermoments,
       author = {{Teague}, Richard and {Foreman-Mackey}, Daniel},
        title = "{A Robust Method to Measure Centroids of Spectral Lines}",
      journal = {Research Notes of the American Astronomical Society},
         year = 2018,
        month = Sep,
       volume = {2},
          eid = {173},
        pages = {173},
          doi = {10.3847/2515-5172/aae265},
       adsurl = {https://ui.adsabs.harvard.edu/abs/2018RNAAS...2c.173T}
}

@ARTICLE{Facchini2024,
       author = {{Facchini}, Stefano and {Testi}, Leonardo and {Humphreys}, Elizabeth and {Vander Donckt}, Mathieu and {Isella}, Andrea and {Wrzosek}, Ramon and {Baudry}, Alain and {Gray}, Malcom D. and {Richards}, Anita M.~S. and {Vlemmmings}, Wouter},
        title = "{Resolved ALMA observations of water in the inner astronomical units of the HL Tau disk}",
      journal = {Nature Astronomy},
         year = 2024,
        month = may,
       volume = {8},
        pages = {587-595},
          doi = {10.1038/s41550-024-02207-w},
archivePrefix = {arXiv},
       eprint = {2403.00647},
 primaryClass = {astro-ph.EP},
       adsurl = {https://ui.adsabs.harvard.edu/abs/2024NatAs...8..587F}
}

@ARTICLE{Bosman2021,
       author = {{Bosman}, Arthur D. and {Bergin}, Edwin A. and {Loomis}, Ryan A. and {Andrews}, Sean M. and {van't Hoff}, Merel L.~R. and {Teague}, Richard and {{\"O}berg}, Karin I. and {Guzm{\'a}n}, Viviana V. and {Walsh}, Catherine and {Aikawa}, Yuri and {Alarc{\'o}n}, Felipe and {Bae}, Jaehan and {Bergner}, Jennifer B. and {Booth}, Alice S. and {Cataldi}, Gianni and {Cleeves}, L. Ilsedore and {Czekala}, Ian and {Huang}, Jane and {Ilee}, John D. and {Law}, Charles J. and {Le Gal}, Romane and {Liu}, Yao and {Long}, Feng and {M{\'e}nard}, Fran{\c{c}}ois and {Nomura}, Hideko and {P{\'e}rez}, Laura M. and {Qi}, Chunhua and {Schwarz}, Kamber R. and {Sierra}, Anibal and {Tsukagoshi}, Takashi and {Yamato}, Yoshihide and {Wilner}, David J. and {Zhang}, Ke},
        title = "{Molecules with ALMA at Planet-forming Scales (MAPS). XV. Tracing Protoplanetary Disk Structure within 20 au}",
      journal = {\apjs},
         year = 2021,
        month = nov,
       volume = {257},
       number = {1},
          eid = {15},
        pages = {15},
          doi = {10.3847/1538-4365/ac1433},
archivePrefix = {arXiv},
       eprint = {2109.06223},
 primaryClass = {astro-ph.EP},
       adsurl = {https://ui.adsabs.harvard.edu/abs/2021ApJS..257...15B}
}

@ARTICLE{Teague2019,
       author = {{Teague}, Richard and {Bae}, Jaehan and {Bergin}, Edwin A.},
        title = "{Meridional flows in the disk around a young star}",
      journal = {\nat},
         year = 2019,
        month = oct,
       volume = {574},
       number = {7778},
        pages = {378-381},
          doi = {10.1038/s41586-019-1642-0},
archivePrefix = {arXiv},
       eprint = {1910.06980},
 primaryClass = {astro-ph.EP},
       adsurl = {https://ui.adsabs.harvard.edu/abs/2019Natur.574..378T}
}

@ARTICLE{Teague2021,
       author = {{Teague}, Richard and {Bae}, Jaehan and {Aikawa}, Yuri and {Andrews}, Sean M. and {Bergin}, Edwin A. and {Bergner}, Jennifer B. and {Boehler}, Yann and {Booth}, Alice S. and {Bosman}, Arthur D. and {Cataldi}, Gianni and {Czekala}, Ian and {Guzm{\'a}n}, Viviana V. and {Huang}, Jane and {Ilee}, John D. and {Law}, Charles J. and {Le Gal}, Romane and {Long}, Feng and {Loomis}, Ryan A. and {M{\'e}nard}, Fran{\c{c}}ois and {{\"O}berg}, Karin I. and {P{\'e}rez}, Laura M. and {Schwarz}, Kamber R. and {Sierra}, Anibal and {Walsh}, Catherine and {Wilner}, David J. and {Yamato}, Yoshihide and {Zhang}, Ke},
        title = "{Molecules with ALMA at Planet-forming Scales (MAPS). XVIII. Kinematic Substructures in the Disks of HD 163296 and MWC 480}",
      journal = {\apjs},
         year = 2021,
        month = nov,
       volume = {257},
       number = {1},
          eid = {18},
        pages = {18},
          doi = {10.3847/1538-4365/ac1438},
archivePrefix = {arXiv},
       eprint = {2109.06218},
 primaryClass = {astro-ph.EP},
       adsurl = {https://ui.adsabs.harvard.edu/abs/2021ApJS..257...18T}
}

@ARTICLE{LAMDA,
       author = {{Sch{\"o}ier}, F.~L. and {van der Tak}, F.~F.~S. and {van Dishoeck}, E.~F. and {Black}, J.~H.},
        title = "{An atomic and molecular database for analysis of submillimetre line observations}",
      journal = {\aap},
         year = 2005,
        month = mar,
       volume = {432},
       number = {1},
        pages = {369-379},
          doi = {10.1051/0004-6361:20041729},
archivePrefix = {arXiv},
       eprint = {astro-ph/0411110},
 primaryClass = {astro-ph},
       adsurl = {https://ui.adsabs.harvard.edu/abs/2005A&A...432..369S}
}

@ARTICLE{Booth2024_HD100546,
       author = {{Booth}, Alice S. and {Leemker}, Margot and {van Dishoeck}, Ewine F. and {Evans}, Lucy and {Ilee}, John D. and {Kama}, Mihkel and {Keyte}, Luke and {Law}, Charles J. and {van der Marel}, Nienke and {Nomura}, Hideko and {Notsu}, Shota and {{\"O}berg}, Karin and {Temmink}, Milou and {Walsh}, Catherine},
        title = "{An ALMA Molecular Inventory of Warm Herbig Ae Disks. I. Molecular Rings, Asymmetries, and Complexity in the HD 100546 Disk}",
      journal = {\aj},
         year = 2024,
        month = apr,
       volume = {167},
       number = {4},
          eid = {164},
        pages = {164},
          doi = {10.3847/1538-3881/ad2700},
archivePrefix = {arXiv},
       eprint = {2402.04001},
 primaryClass = {astro-ph.EP},
       adsurl = {https://ui.adsabs.harvard.edu/abs/2024AJ....167..164B}
}

@ARTICLE{Booth2024_IRS48,
       author = {{Booth}, Alice S. and {Temmink}, Milou and {van Dishoeck}, Ewine F. and {Evans}, Lucy and {Ilee}, John D. and {Kama}, Mihkel and {Keyte}, Luke and {Law}, Charles J. and {Leemker}, Margot and {van der Marel}, Nienke and {Nomura}, Hideko and {Notsu}, Shota and {{\"O}berg}, Karin and {Walsh}, Catherine},
        title = "{An ALMA Molecular Inventory of Warm Herbig Ae Disks. II. Abundant Complex Organics and Volatile Sulphur in the IRS 48 Disk}",
      journal = {\aj},
         year = 2024,
        month = apr,
       volume = {167},
       number = {4},
          eid = {165},
        pages = {165},
          doi = {10.3847/1538-3881/ad26ff},
archivePrefix = {arXiv},
       eprint = {2402.04002},
 primaryClass = {astro-ph.EP},
       adsurl = {https://ui.adsabs.harvard.edu/abs/2024AJ....167..165B}
}

@ARTICLE{Semenov2018,
       author = {{Semenov}, D. and {Favre}, C. and {Fedele}, D. and {Guilloteau}, S. and {Teague}, R. and {Henning}, Th. and {Dutrey}, A. and {Chapillon}, E. and {Hersant}, F. and {Pi{\'e}tu}, V.},
        title = "{Chemistry in disks. XI. Sulfur-bearing species as tracers of protoplanetary disk physics and chemistry: the DM Tau case}",
      journal = {\aap},
         year = 2018,
        month = sep,
       volume = {617},
          eid = {A28},
        pages = {A28},
          doi = {10.1051/0004-6361/201832980},
archivePrefix = {arXiv},
       eprint = {1806.07707},
 primaryClass = {astro-ph.GA},
       adsurl = {https://ui.adsabs.harvard.edu/abs/2018A&A...617A..28S}
}

@ARTICLE{emcee,
       author = {{Foreman-Mackey}, Daniel and {Hogg}, David W. and {Lang}, Dustin and {Goodman}, Jonathan},
        title = "{emcee: The MCMC Hammer}",
      journal = {\pasp},
         year = 2013,
        month = mar,
       volume = {125},
       number = {925},
        pages = {306},
          doi = {10.1086/670067},
archivePrefix = {arXiv},
       eprint = {1202.3665},
 primaryClass = {astro-ph.IM},
       adsurl = {https://ui.adsabs.harvard.edu/abs/2013PASP..125..306F}
}

@ARTICLE{Loomis2018,
       author = {{Loomis}, Ryan A. and {{\"O}berg}, Karin I. and {Andrews}, Sean M. and {Walsh}, Catherine and {Czekala}, Ian and {Huang}, Jane and {Rosenfeld}, Katherine A.},
        title = "{Detecting Weak Spectral Lines in Interferometric Data through Matched Filtering}",
      journal = {\aj},
         year = 2018,
        month = apr,
       volume = {155},
       number = {4},
          eid = {182},
        pages = {182},
          doi = {10.3847/1538-3881/aab604},
archivePrefix = {arXiv},
       eprint = {1803.04987},
 primaryClass = {astro-ph.IM},
       adsurl = {https://ui.adsabs.harvard.edu/abs/2018AJ....155..182L}
}

@ARTICLE{Notsu2019,
       author = {{Notsu}, Shota and {Akiyama}, Eiji and {Booth}, Alice and {Nomura}, Hideko and {Walsh}, Catherine and {Hirota}, Tomoya and {Honda}, Mitsuhiko and {Tsukagoshi}, Takashi and {Millar}, T.~J.},
        title = "{Dust Continuum Emission and the Upper Limit Fluxes of Submillimeter Water Lines of the Protoplanetary Disk around HD 163296 Observed by ALMA}",
      journal = {\apj},
         year = 2019,
        month = apr,
       volume = {875},
       number = {2},
          eid = {96},
        pages = {96},
          doi = {10.3847/1538-4357/ab0ae9},
archivePrefix = {arXiv},
       eprint = {1902.09932},
 primaryClass = {astro-ph.EP},
       adsurl = {https://ui.adsabs.harvard.edu/abs/2019ApJ...875...96N}
}

@ARTICLE{Wilson1999,
       author = {{Wilson}, T.~L.},
        title = "{Isotopes in the interstellar medium and circumstellar envelopes}",
      journal = {Reports on Progress in Physics},
         year = 1999,
        month = feb,
       volume = {62},
       number = {2},
        pages = {143-185},
          doi = {10.1088/0034-4885/62/2/002},
       adsurl = {https://ui.adsabs.harvard.edu/abs/1999RPPh...62..143W}
}

@ARTICLE{Andreu2023,
       author = {{Andreu}, A. and {Coutens}, A. and {Cruz-S{\'a}enz de Miera}, F. and {Houry}, N. and {J{\o}rgensen}, J.~K. and {K{\'o}sp{\'a}l}, A. and {Harsono}, D.},
        title = "{A high HDO/H$_{2}$O ratio in the Class I protostar L1551 IRS5}",
      journal = {\aap},
         year = 2023,
        month = sep,
       volume = {677},
          eid = {L17},
        pages = {L17},
          doi = {10.1051/0004-6361/202347484},
archivePrefix = {arXiv},
       eprint = {2309.01688},
 primaryClass = {astro-ph.SR},
       adsurl = {https://ui.adsabs.harvard.edu/abs/2023A&A...677L..17A}
}

@ARTICLE{Tobin2023,
       author = {{Tobin}, John J. and {van't Hoff}, Merel L.~R. and {Leemker}, Margot and {van Dishoeck}, Ewine F. and {Paneque-Carre{\~n}o}, Teresa and {Furuya}, Kenji and {Harsono}, Daniel and {Persson}, Magnus V. and {Cleeves}, L. Ilsedore and {Sheehan}, Patrick D. and {Cieza}, Lucas},
        title = "{Deuterium-enriched water ties planet-forming disks to comets and protostars}",
      journal = {\nat},
         year = 2023,
        month = mar,
       volume = {615},
       number = {7951},
        pages = {227-230},
          doi = {10.1038/s41586-022-05676-z},
       adsurl = {https://ui.adsabs.harvard.edu/abs/2023Natur.615..227T}
}

@ARTICLE{McClure2023,
       author = {{McClure}, M.~K. and {Rocha}, W.~R.~M. and {Pontoppidan}, K.~M. and {Crouzet}, N. and {Chu}, L.~E.~U. and {Dartois}, E. and {Lamberts}, T. and {Noble}, J.~A. and {Pendleton}, Y.~J. and {Perotti}, G. and {Qasim}, D. and {Rachid}, M.~G. and {Smith}, Z.~L. and {Sun}, Fengwu and {Beck}, Tracy L. and {Boogert}, A.~C.~A. and {Brown}, W.~A. and {Caselli}, P. and {Charnley}, S.~B. and {Cuppen}, Herma M. and {Dickinson}, H. and {Drozdovskaya}, M.~N. and {Egami}, E. and {Erkal}, J. and {Fraser}, H. and {Garrod}, R.~T. and {Harsono}, D. and {Ioppolo}, S. and {Jim{\'e}nez-Serra}, I. and {Jin}, M. and {J{\o}rgensen}, J.~K. and {Kristensen}, L.~E. and {Lis}, D.~C. and {McCoustra}, M.~R.~S. and {McGuire}, Brett A. and {Melnick}, G.~J. and {{\"O}berg}, Karin I. and {Palumbo}, M.~E. and {Shimonishi}, T. and {Sturm}, J.~A. and {van Dishoeck}, E.~F. and {Linnartz}, H.},
        title = "{An Ice Age JWST inventory of dense molecular cloud ices}",
      journal = {Nature Astronomy},
         year = 2023,
        month = apr,
       volume = {7},
        pages = {431-443},
          doi = {10.1038/s41550-022-01875-w},
archivePrefix = {arXiv},
       eprint = {2301.09140},
 primaryClass = {astro-ph.GA},
       adsurl = {https://ui.adsabs.harvard.edu/abs/2023NatAs...7..431M}
}

@ARTICLE{Phuong2018,
       author = {{Phuong}, N.~T. and {Chapillon}, E. and {Majumdar}, L. and {Dutrey}, A. and {Guilloteau}, S. and {Pi{\'e}tu}, V. and {Wakelam}, V. and {Diep}, P.~N. and {Tang}, Y. -W. and {Beck}, T. and {Bary}, J.},
        title = "{First detection of H$_{2}$S in a protoplanetary disk. The dense GG Tauri A ring}",
      journal = {\aap},
         year = 2018,
        month = aug,
       volume = {616},
          eid = {L5},
        pages = {L5},
          doi = {10.1051/0004-6361/201833766},
archivePrefix = {arXiv},
       eprint = {1808.00652},
 primaryClass = {astro-ph.SR},
       adsurl = {https://ui.adsabs.harvard.edu/abs/2018A&A...616L...5P}
}

@ARTICLE{Riviere-Marichalar2022,
       author = {{Rivi{\`e}re-Marichalar}, P. and {Fuente}, A. and {Esplugues}, G. and {Wakelam}, V. and {le Gal}, R. and {Baruteau}, C. and {Ribas}, A. and {Mac{\'\i}as}, E. and {Neri}, R. and {Navarro-Almaida}, D.},
        title = "{AB Aur, a Rosetta stone for studies of planet formation. II. H$_{2}$S detection and sulfur budget}",
      journal = {\aap},
         year = 2022,
        month = sep,
       volume = {665},
          eid = {A61},
        pages = {A61},
          doi = {10.1051/0004-6361/202142906},
archivePrefix = {arXiv},
       eprint = {2207.06716},
 primaryClass = {astro-ph.SR},
       adsurl = {https://ui.adsabs.harvard.edu/abs/2022A&A...665A..61R}
}

@ARTICLE{Riviere-Marichalar2021,
       author = {{Rivi{\`e}re-Marichalar}, P. and {Fuente}, A. and {Le Gal}, R. and {Arabhavi}, A.~M. and {Cazaux}, S. and {Navarro-Almaida}, D. and {Ribas}, A. and {Mendigut{\'\i}a}, I. and {Barrado}, D. and {Montesinos}, B.},
        title = "{H$_{2}$S observations in young stellar disks in Taurus}",
      journal = {\aap},
         year = 2021,
        month = aug,
       volume = {652},
          eid = {A46},
        pages = {A46},
          doi = {10.1051/0004-6361/202140470},
archivePrefix = {arXiv},
       eprint = {2106.02430},
 primaryClass = {astro-ph.SR},
       adsurl = {https://ui.adsabs.harvard.edu/abs/2021A&A...652A..46R}
}

@ARTICLE{Santos2025,
       author = {{Santos}, Julia C. and {Piacentino}, Elettra L. and {Bergner}, Jennifer B. and {Rajappan}, Mahesh and {{\"O}berg}, Karin I.},
        title = "{H2S ice sublimation dynamics: experimentally constrained binding energies, entrapment efficiencies, and snowlines}",
      journal = {arXiv e-prints},
         year = 2025,
        month = apr,
          eid = {arXiv:2504.14010},
        pages = {arXiv:2504.14010},
          doi = {10.48550/arXiv.2504.14010},
archivePrefix = {arXiv},
       eprint = {2504.14010},
 primaryClass = {astro-ph.EP},
       adsurl = {https://ui.adsabs.harvard.edu/abs/2025arXiv250414010S}
}

@ARTICLE{LeGal2019,
       author = {{Le Gal}, Romane and {{\"O}berg}, Karin I. and {Loomis}, Ryan A. and {Pegues}, Jamila and {Bergner}, Jennifer B.},
        title = "{Sulfur Chemistry in Protoplanetary Disks: CS and H$_{2}$CS}",
      journal = {\apj},
         year = 2019,
        month = may,
       volume = {876},
       number = {1},
          eid = {72},
        pages = {72},
          doi = {10.3847/1538-4357/ab1416},
archivePrefix = {arXiv},
       eprint = {1903.11105},
 primaryClass = {astro-ph.GA},
       adsurl = {https://ui.adsabs.harvard.edu/abs/2019ApJ...876...72L}
}

@ARTICLE{LeGal2021,
       author = {{Le Gal}, Romane and {{\"O}berg}, Karin I. and {Teague}, Richard and {Loomis}, Ryan A. and {Law}, Charles J. and {Walsh}, Catherine and {Bergin}, Edwin A. and {M{\'e}nard}, Fran{\c{c}}ois and {Wilner}, David J. and {Andrews}, Sean M. and {Aikawa}, Yuri and {Booth}, Alice S. and {Cataldi}, Gianni and {Bergner}, Jennifer B. and {Bosman}, Arthur D. and {Cleeves}, L. Ilse and {Czekala}, Ian and {Furuya}, Kenji and {Guzm{\'a}n}, Viviana V. and {Huang}, Jane and {Ilee}, John D. and {Nomura}, Hideko and {Qi}, Chunhua and {Schwarz}, Kamber R. and {Tsukagoshi}, Takashi and {Yamato}, Yoshihide and {Zhang}, Ke},
        title = "{Molecules with ALMA at Planet-forming Scales (MAPS). XII. Inferring the C/O and S/H Ratios in Protoplanetary Disks with Sulfur Molecules}",
      journal = {\apjs},
         year = 2021,
        month = nov,
       volume = {257},
       number = {1},
          eid = {12},
        pages = {12},
          doi = {10.3847/1538-4365/ac2583},
archivePrefix = {arXiv},
       eprint = {2109.06286},
 primaryClass = {astro-ph.GA},
       adsurl = {https://ui.adsabs.harvard.edu/abs/2021ApJS..257...12L}
}

@ARTICLE{Dutrey1997,
       author = {{Dutrey}, A. and {Guilloteau}, S. and {Guelin}, M.},
        title = "{Chemistry of protosolar-like nebulae: The molecular content of the DM Tau and GG Tau disks.}",
      journal = {\aap},
         year = 1997,
        month = jan,
       volume = {317},
        pages = {L55-L58},
       adsurl = {https://ui.adsabs.harvard.edu/abs/1997A&A...317L..55D}
}

@ARTICLE{Guilloteau2016,
       author = {{Guilloteau}, S. and {Reboussin}, L. and {Dutrey}, A. and {Chapillon}, E. and {Wakelam}, V. and {Pi{\'e}tu}, V. and {Di Folco}, E. and {Semenov}, D. and {Henning}, Th.},
        title = "{Chemistry in disks. X. The molecular content of protoplanetary disks in Taurus}",
      journal = {\aap},
         year = 2016,
        month = aug,
       volume = {592},
          eid = {A124},
        pages = {A124},
          doi = {10.1051/0004-6361/201527088},
archivePrefix = {arXiv},
       eprint = {1604.05028},
 primaryClass = {astro-ph.EP},
       adsurl = {https://ui.adsabs.harvard.edu/abs/2016A&A...592A.124G}
}

@ARTICLE{Booth2023,
       author = {{Booth}, Alice S. and {Ilee}, John D. and {Walsh}, Catherine and {Kama}, Mihkel and {Keyte}, Luke and {van Dishoeck}, Ewine F. and {Nomura}, Hideko},
        title = "{Sulphur monoxide emission tracing an embedded planet in the HD 100546 protoplanetary disk}",
      journal = {\aap},
         year = 2023,
        month = jan,
       volume = {669},
          eid = {A53},
        pages = {A53},
          doi = {10.1051/0004-6361/202244472},
archivePrefix = {arXiv},
       eprint = {2210.14820},
 primaryClass = {astro-ph.EP},
       adsurl = {https://ui.adsabs.harvard.edu/abs/2023A&A...669A..53B}
}

@ARTICLE{Booth2023_HD169142,
       author = {{Booth}, Alice S. and {Law}, Charles J. and {Temmink}, Milou and {Leemker}, Margot and {Mac{\'\i}as}, Enrique},
        title = "{Tracing snowlines and C/O ratio in a planet-hosting disk. ALMA molecular line observations towards the HD 169142 disk}",
      journal = {\aap},
         year = 2023,
        month = oct,
       volume = {678},
          eid = {A146},
        pages = {A146},
          doi = {10.1051/0004-6361/202346974},
archivePrefix = {arXiv},
       eprint = {2308.07910},
 primaryClass = {astro-ph.EP},
       adsurl = {https://ui.adsabs.harvard.edu/abs/2023A&A...678A.146B}
}

@ARTICLE{Booth2021,
       author = {{Booth}, Alice S. and {van der Marel}, Nienke and {Leemker}, Margot and {van Dishoeck}, Ewine F. and {Ohashi}, Satoshi},
        title = "{A major asymmetric ice trap in a planet-forming disk. II. Prominent SO and SO$_{2}$ pointing to C/O < 1}",
      journal = {\aap},
         year = 2021,
        month = jul,
       volume = {651},
          eid = {L6},
        pages = {L6},
          doi = {10.1051/0004-6361/202141057},
archivePrefix = {arXiv},
       eprint = {2104.08908},
 primaryClass = {astro-ph.EP},
       adsurl = {https://ui.adsabs.harvard.edu/abs/2021A&A...651L...6B}
}

@ARTICLE{Keyte2024,
       author = {{Keyte}, Luke and {Kama}, Mihkel and {Chuang}, Ko-Ju and {Cleeves}, L. Ilsedore and {Drozdovskaya}, Maria N. and {Furuya}, Kenji and {Rawlings}, Jonathan and {Shorttle}, Oliver},
        title = "{Spatially resolving the volatile sulfur abundance in the HD 100546 protoplanetary disc}",
      journal = {\mnras},
         year = 2024,
        month = feb,
       volume = {528},
       number = {1},
        pages = {388-407},
          doi = {10.1093/mnras/stae019},
archivePrefix = {arXiv},
       eprint = {2312.13997},
 primaryClass = {astro-ph.EP},
       adsurl = {https://ui.adsabs.harvard.edu/abs/2024MNRAS.528..388K}
}

@ARTICLE{Dutrey2011,
       author = {{Dutrey}, A. and {Wakelam}, V. and {Boehler}, Y. and {Guilloteau}, S. and {Hersant}, F. and {Semenov}, D. and {Chapillon}, E. and {Henning}, T. and {Pi{\'e}tu}, V. and {Launhardt}, R. and {Gueth}, F. and {Schreyer}, K.},
        title = "{Chemistry in disks. V. Sulfur-bearing molecules in the protoplanetary disks surrounding LkCa15, MWC480, DM Tauri, and GO Tauri}",
      journal = {\aap},
         year = 2011,
        month = nov,
       volume = {535},
          eid = {A104},
        pages = {A104},
          doi = {10.1051/0004-6361/201116931},
archivePrefix = {arXiv},
       eprint = {1109.5870},
 primaryClass = {astro-ph.SR},
       adsurl = {https://ui.adsabs.harvard.edu/abs/2011A&A...535A.104D}
}

@ARTICLE{Law2025,
       author = {{Law}, Charles J. and {Le Gal}, Romane and {Yamato}, Yoshihide and {Zhang}, Ke and {Guzm{\'a}n}, Viviana V. and {Hern{\'a}ndez-Vera}, Claudio and {Cleeves}, L. Ilsedore and {Guidi}, Greta and {Booth}, Alice S.},
        title = "{A Multiline Analysis of the Distribution and Excitation of CS and H$_{2}$CS in the HD 163296 Disk}",
      journal = {\apj},
         year = 2025,
        month = may,
       volume = {985},
       number = {1},
          eid = {84},
        pages = {84},
          doi = {10.3847/1538-4357/adc304},
archivePrefix = {arXiv},
       eprint = {2503.16605},
 primaryClass = {astro-ph.EP},
       adsurl = {https://ui.adsabs.harvard.edu/abs/2025ApJ...985...84L}
}

@ARTICLE{Calmonte2016,
       author = {{Calmonte}, U. and {Altwegg}, K. and {Balsiger}, H. and {Berthelier}, J.~J. and {Bieler}, A. and {Cessateur}, G. and {Dhooghe}, F. and {van Dishoeck}, E.~F. and {Fiethe}, B. and {Fuselier}, S.~A. and {Gasc}, S. and {Gombosi}, T.~I. and {H{\"a}ssig}, M. and {Le Roy}, L. and {Rubin}, M. and {S{\'e}mon}, T. and {Tzou}, C. -Y. and {Wampfler}, S.~F.},
        title = "{Sulphur-bearing species in the coma of comet 67P/Churyumov-Gerasimenko}",
      journal = {\mnras},
         year = 2016,
        month = nov,
       volume = {462},
        pages = {S253-S273},
          doi = {10.1093/mnras/stw2601},
       adsurl = {https://ui.adsabs.harvard.edu/abs/2016MNRAS.462S.253C}
}

@ARTICLE{Palumdo1997,
       author = {{Palumbo}, M.~E. and {Geballe}, T.~R. and {Tielens}, A.~G.~G.~M.},
        title = "{Solid Carbonyl Sulfide (OCS) in Dense Molecular Clouds}",
      journal = {\apj},
         year = 1997,
        month = apr,
       volume = {479},
       number = {2},
        pages = {839-844},
          doi = {10.1086/303905},
       adsurl = {https://ui.adsabs.harvard.edu/abs/1997ApJ...479..839P}
}

@ARTICLE{Palumdo1995,
       author = {{Palumbo}, M.~E. and {Tielens}, A.~G.~G.~M. and {Tokunaga}, A.~T.},
        title = "{Solid Carbonyl Sulphide (OCS) in W33A}",
      journal = {\apj},
         year = 1995,
        month = aug,
       volume = {449},
        pages = {674},
          doi = {10.1086/176088},
       adsurl = {https://ui.adsabs.harvard.edu/abs/1995ApJ...449..674P}
}

@ARTICLE{Aikawa2012,
       author = {{Aikawa}, Y. and {Kamuro}, D. and {Sakon}, I. and {Itoh}, Y. and {Terada}, H. and {Noble}, J.~A. and {Pontoppidan}, K.~M. and {Fraser}, H.~J. and {Tamura}, M. and {Kandori}, R. and {Kawamura}, A. and {Ueno}, M.},
        title = "{AKARI observations of ice absorption bands towards edge-on young stellar objects}",
      journal = {\aap},
         year = 2012,
        month = feb,
       volume = {538},
          eid = {A57},
        pages = {A57},
          doi = {10.1051/0004-6361/201015999},
       adsurl = {https://ui.adsabs.harvard.edu/abs/2012A&A...538A..57A}
}

@ARTICLE{Kama2019,
       author = {{Kama}, Mihkel and {Shorttle}, Oliver and {Jermyn}, Adam S. and {Folsom}, Colin P. and {Furuya}, Kenji and {Bergin}, Edwin A. and {Walsh}, Catherine and {Keller}, Lindsay},
        title = "{Abundant Refractory Sulfur in Protoplanetary Disks}",
      journal = {\apj},
         year = 2019,
        month = nov,
       volume = {885},
       number = {2},
          eid = {114},
        pages = {114},
          doi = {10.3847/1538-4357/ab45f8},
archivePrefix = {arXiv},
       eprint = {1908.05169},
 primaryClass = {astro-ph.EP},
       adsurl = {https://ui.adsabs.harvard.edu/abs/2019ApJ...885..114K}
}

@ARTICLE{Drozdovskaya2018,
       author = {{Drozdovskaya}, Maria N. and {van Dishoeck}, Ewine F. and {J{\o}rgensen}, Jes K. and {Calmonte}, Ursina and {van der Wiel}, Matthijs H.~D. and {Coutens}, Audrey and {Calcutt}, Hannah and {M{\"u}ller}, Holger S.~P. and {Bjerkeli}, Per and {Persson}, Magnus V. and {Wampfler}, Susanne F. and {Altwegg}, Kathrin},
        title = "{The ALMA-PILS survey: the sulphur connection between protostars and comets: IRAS 16293-2422 B and 67P/Churyumov-Gerasimenko}",
      journal = {\mnras},
         year = 2018,
        month = jun,
       volume = {476},
       number = {4},
        pages = {4949-4964},
          doi = {10.1093/mnras/sty462},
archivePrefix = {arXiv},
       eprint = {1802.02977},
 primaryClass = {astro-ph.SR},
       adsurl = {https://ui.adsabs.harvard.edu/abs/2018MNRAS.476.4949D}
}

@ARTICLE{Kushwahaa2023,
       author = {{Kushwahaa}, Tanya and {Drozdovskaya}, Maria N. and {Tychoniec}, {\L}ukasz and {Tabone}, Beno{\^\i}t},
        title = "{ALMA ACA study of the H$_{2}$S/OCS ratio in low-mass protostars}",
      journal = {\aap},
         year = 2023,
        month = apr,
       volume = {672},
          eid = {A122},
        pages = {A122},
          doi = {10.1051/0004-6361/202245097},
archivePrefix = {arXiv},
       eprint = {2302.09452},
 primaryClass = {astro-ph.GA},
       adsurl = {https://ui.adsabs.harvard.edu/abs/2023A&A...672A.122K}
}

@ARTICLE{Miranzo-Pastor2025,
       author = {{Miranzo-Pastor}, J.~J. and {Fuente}, A. and {Navarro-Almaida}, D. and {Pineda}, J.~E. and {Segura-Cox}, D.~M. and {Caselli}, P. and {Martin-Domenech}, R. and {Valdivia-Mena}, M.~T. and {Henning}, T. and {Hsieh}, T. -H. and {Busch}, L.~A. and {Gieser}, C. and {Chou}, Y. -R. and {Commer{\c{c}}on}, B. and {Neri}, R. and {Semenov}, D. and {Lopez-Sepulcre}, A. and {Cunningham}, N. and {Bouscasse}, L. and {Maureira}, M.},
        title = "{PRODIGE VI -- Envelope to Disk with NOEMA: VI. The Missing Sulfur Problem}",
      journal = {arXiv e-prints},
         year = 2025,
        month = jul,
          eid = {arXiv:2507.05830},
        pages = {arXiv:2507.05830},
          doi = {10.48550/arXiv.2507.05830},
archivePrefix = {arXiv},
       eprint = {2507.05830},
 primaryClass = {astro-ph.GA},
       adsurl = {https://ui.adsabs.harvard.edu/abs/2025arXiv250705830M}
}

@ARTICLE{Bockelee-Morvan2000,
       author = {{Bockel{\'e}e-Morvan}, D. and {Lis}, D.~C. and {Wink}, J.~E. and {Despois}, D. and {Crovisier}, J. and {Bachiller}, R. and {Benford}, D.~J. and {Biver}, N. and {Colom}, P. and {Davies}, J.~K. and {G{\'e}rard}, E. and {Germain}, B. and {Houde}, M. and {Mehringer}, D. and {Moreno}, R. and {Paubert}, G. and {Phillips}, T.~G. and {Rauer}, H.},
        title = "{New molecules found in comet C/1995 O1 (Hale-Bopp). Investigating the link between cometary and interstellar material}",
      journal = {\aap},
         year = 2000,
        month = jan,
       volume = {353},
        pages = {1101-1114},
       adsurl = {https://ui.adsabs.harvard.edu/abs/2000A&A...353.1101B}
}

@ARTICLE{Booth2018,
       author = {{Booth}, Alice S. and {Walsh}, Catherine and {Kama}, Mihkel and {Loomis}, Ryan A. and {Maud}, Luke T. and {Juh{\'a}sz}, Attila},
        title = "{Sulphur monoxide exposes a potential molecular disk wind from the planet-hosting disk around HD 100546}",
      journal = {\aap},
         year = 2018,
        month = mar,
       volume = {611},
          eid = {A16},
        pages = {A16},
          doi = {10.1051/0004-6361/201731347},
archivePrefix = {arXiv},
       eprint = {1712.05992},
 primaryClass = {astro-ph.EP},
       adsurl = {https://ui.adsabs.harvard.edu/abs/2018A&A...611A..16B}
}

@ARTICLE{Asplund2020,
       author = {{Asplund}, M. and {Amarsi}, A.~M. and {Grevesse}, N.},
        title = "{The chemical make-up of the Sun: A 2020 vision}",
      journal = {\aap},
         year = 2021,
        month = sep,
       volume = {653},
          eid = {A141},
        pages = {A141},
          doi = {10.1051/0004-6361/202140445},
archivePrefix = {arXiv},
       eprint = {2105.01661},
 primaryClass = {astro-ph.SR},
       adsurl = {https://ui.adsabs.harvard.edu/abs/2021A&A...653A.141A}
}

@ARTICLE{Ranjan2018,
       author = {{Ranjan}, Sukrit and {Todd}, Zoe R. and {Sutherland}, John D. and {Sasselov}, Dimitar D.},
        title = "{Sulfidic Anion Concentrations on Early Earth for Surficial Origins-of-Life Chemistry}",
      journal = {Astrobiology},
         year = 2018,
        month = aug,
       volume = {18},
       number = {8},
        pages = {1023-1040},
          doi = {10.1089/ast.2017.1770},
archivePrefix = {arXiv},
       eprint = {1801.07725},
 primaryClass = {astro-ph.EP},
       adsurl = {https://ui.adsabs.harvard.edu/abs/2018AsBio..18.1023R}
}

@ARTICLE{Tieftruck1994,
       author = {{Tieftrunk}, A. and {Pineau des Forets}, G. and {Schilke}, P. and {Walmsley}, C.~M.},
        title = "{SO and H\_2\_S in low density molecular clouds.}",
      journal = {\aap},
         year = 1994,
        month = sep,
       volume = {289},
        pages = {579-596},
       adsurl = {https://ui.adsabs.harvard.edu/abs/1994A&A...289..579T}
}

@ARTICLE{Boogert1997,
       author = {{Boogert}, A.~C.~A. and {Schutte}, W.~A. and {Helmich}, F.~P. and {Tielens}, A.~G.~G.~M. and {Wooden}, D.~H.},
        title = "{Infrared observations and laboratory simulations of interstellar CH\_4\_ and SO\_2\_.}",
      journal = {\aap},
         year = 1997,
        month = feb,
       volume = {317},
        pages = {929-941},
       adsurl = {https://ui.adsabs.harvard.edu/abs/1997A&A...317..929B}
}

@ARTICLE{Boogert2015,
       author = {{Boogert}, A.~C. Adwin and {Gerakines}, Perry A. and {Whittet}, Douglas C.~B.},
        title = "{Observations of the icy universe.}",
      journal = {\araa},
         year = 2015,
        month = aug,
       volume = {53},
        pages = {541-581},
          doi = {10.1146/annurev-astro-082214-122348},
archivePrefix = {arXiv},
       eprint = {1501.05317},
 primaryClass = {astro-ph.GA},
       adsurl = {https://ui.adsabs.harvard.edu/abs/2015ARA&A..53..541B}
}

@ARTICLE{Rocha2024,
       author = {{Rocha}, W.~R.~M. and {van Dishoeck}, E.~F. and {Ressler}, M.~E. and {van Gelder}, M.~L. and {Slavicinska}, K. and {Brunken}, N.~G.~C. and {Linnartz}, H. and {Ray}, T.~P. and {Beuther}, H. and {Caratti o Garatti}, A. and {Geers}, V. and {Kavanagh}, P.~J. and {Klaassen}, P.~D. and {Justtanont}, K. and {Chen}, Y. and {Francis}, L. and {Gieser}, C. and {Perotti}, G. and {Tychoniec}, {\L}. and {Barsony}, M. and {Majumdar}, L. and {le Gouellec}, V.~J.~M. and {Chu}, L.~E.~U. and {Lew}, B.~W.~P. and {Henning}, Th. and {Wright}, G.},
        title = "{JWST Observations of Young protoStars (JOYS+): Detecting icy complex organic molecules and ions. I. CH$_{4}$, SO$_{2}$, HCOO$^{{\ensuremath{-}}}$, OCN$^{{\ensuremath{-}}}$, H$_{2}$CO, HCOOH, CH$_{3}$CH$_{2}$OH, CH$_{3}$CHO, CH$_{3}$OCHO, and CH$_{3}$COOH}",
      journal = {\aap},
         year = 2024,
        month = mar,
       volume = {683},
          eid = {A124},
        pages = {A124},
          doi = {10.1051/0004-6361/202348427},
archivePrefix = {arXiv},
       eprint = {2312.06834},
 primaryClass = {astro-ph.SR},
       adsurl = {https://ui.adsabs.harvard.edu/abs/2024A&A...683A.124R}
}

@ARTICLE{Slavicinska2025,
       author = {{Slavicinska}, K. and {Boogert}, A.~C.~A. and {Tychoniec}, {\L}. and {van Dishoeck}, E.~F. and {van Gelder}, M.~L. and {Navarro}, M.~G. and {Santos}, J.~C. and {Klaassen}, P.~D. and {Kavanagh}, P.~J. and {Chuang}, K.-J.},
        title = "{Ammonium hydrosulfide (NH$_{4}$SH) as a potentially significant sulfur sink in interstellar ices}",
      journal = {\aap},
         year = 2025,
        month = jan,
       volume = {693},
          eid = {A146},
        pages = {A146},
          doi = {10.1051/0004-6361/202451383},
archivePrefix = {arXiv},
       eprint = {2410.02860},
 primaryClass = {astro-ph.GA},
       adsurl = {https://ui.adsabs.harvard.edu/abs/2025A&A...693A.146S}
}

@ARTICLE{Gaia2023,
       author = {{Gaia Collaboration} and {Vallenari}, A. and {Brown}, A.~G.~A. and {Prusti}, T. and {de Bruijne}, J.~H.~J. and {Arenou}, F. and {Babusiaux}, C. and {Biermann}, M. and {Creevey}, O.~L. and {Ducourant}, C. and {Evans}, D.~W. and {Eyer}, L. and {Guerra}, R. and {Hutton}, A. and {Jordi}, C. and {Klioner}, S.~A. and {Lammers}, U.~L. and {Lindegren}, L. and {Luri}, X. and {Mignard}, F. and {Panem}, C. and {Pourbaix}, D. and {Randich}, S. and {Sartoretti}, P. and {Soubiran}, C. and {Tanga}, P. and {Walton}, N.~A. and {Bailer-Jones}, C.~A.~L. and {Bastian}, U. and {Drimmel}, R. and {Jansen}, F. and {Katz}, D. and {Lattanzi}, M.~G. and {van Leeuwen}, F. and {Bakker}, J. and {Cacciari}, C. and {Casta{\~n}eda}, J. and {De Angeli}, F. and {Fabricius}, C. and {Fouesneau}, M. and {Fr{\'e}mat}, Y. and {Galluccio}, L. and {Guerrier}, A. and {Heiter}, U. and {Masana}, E. and {Messineo}, R. and {Mowlavi}, N. and {Nicolas}, C. and {Nienartowicz}, K. and {Pailler}, F. and {Panuzzo}, P. and {Riclet}, F. and {Roux}, W. and {Seabroke}, G.~M. and {Sordo}, R. and {Th{\'e}venin}, F. and {Gracia-Abril}, G. and {Portell}, J. and {Teyssier}, D. and {Altmann}, M. and {Andrae}, R. and {Audard}, M. and {Bellas-Velidis}, I. and {Benson}, K. and {Berthier}, J. and {Blomme}, R. and {Burgess}, P.~W. and {Busonero}, D. and {Busso}, G. and {C{\'a}novas}, H. and {Carry}, B. and {Cellino}, A. and {Cheek}, N. and {Clementini}, G. and {Damerdji}, Y. and {Davidson}, M. and {de Teodoro}, P. and {Nu{\~n}ez Campos}, M. and {Delchambre}, L. and {Dell'Oro}, A. and {Esquej}, P. and {Fern{\'a}ndez-Hern{\'a}ndez}, J. and {Fraile}, E. and {Garabato}, D. and {Garc{\'\i}a-Lario}, P. and {Gosset}, E. and {Haigron}, R. and {Halbwachs}, J.-L. and {Hambly}, N.~C. and {Harrison}, D.~L. and {Hern{\'a}ndez}, J. and {Hestroffer}, D. and {Hodgkin}, S.~T. and {Holl}, B. and {Jan{\ss}en}, K. and {Jevardat de Fombelle}, G. and {Jordan}, S. and {Krone-Martins}, A. and {Lanzafame}, A.~C. and {L{\"o}ffler}, W. and {Marchal}, O. and {Marrese}, P.~M. and {Moitinho}, A. and {Muinonen}, K. and {Osborne}, P. and {Pancino}, E. and {Pauwels}, T. and {Recio-Blanco}, A. and {Reyl{\'e}}, C. and {Riello}, M. and {Rimoldini}, L. and {Roegiers}, T. and {Rybizki}, J. and {Sarro}, L.~M. and {Siopis}, C. and {Smith}, M. and {Sozzetti}, A. and {Utrilla}, E. and {van Leeuwen}, M. and {Abbas}, U. and {{\'A}brah{\'a}m}, P. and {Abreu Aramburu}, A. and {Aerts}, C. and {Aguado}, J.~J. and {Ajaj}, M. and {Aldea-Montero}, F. and {Altavilla}, G. and {{\'A}lvarez}, M.~A. and {Alves}, J. and {Anders}, F. and {Anderson}, R.~I. and {Anglada Varela}, E. and {Antoja}, T. and {Baines}, D. and {Baker}, S.~G. and {Balaguer-N{\'u}{\~n}ez}, L. and {Balbinot}, E. and {Balog}, Z. and {Barache}, C. and {Barbato}, D. and {Barros}, M. and {Barstow}, M.~A. and {Bartolom{\'e}}, S. and {Bassilana}, J.-L. and {Bauchet}, N. and {Becciani}, U. and {Bellazzini}, M. and {Berihuete}, A. and {Bernet}, M. and {Bertone}, S. and {Bianchi}, L. and {Binnenfeld}, A. and {Blanco-Cuaresma}, S. and {Blazere}, A. and {Boch}, T. and {Bombrun}, A. and {Bossini}, D. and {Bouquillon}, S. and {Bragaglia}, A. and {Bramante}, L. and {Breedt}, E. and {Bressan}, A. and {Brouillet}, N. and {Brugaletta}, E. and {Bucciarelli}, B. and {Burlacu}, A. and {Butkevich}, A.~G. and {Buzzi}, R. and {Caffau}, E. and {Cancelliere}, R. and {Cantat-Gaudin}, T. and {Carballo}, R. and {Carlucci}, T. and {Carnerero}, M.~I. and {Carrasco}, J.~M. and {Casamiquela}, L. and {Castellani}, M. and {Castro-Ginard}, A. and {Chaoul}, L. and {Charlot}, P. and {Chemin}, L. and {Chiaramida}, V. and {Chiavassa}, A. and {Chornay}, N. and {Comoretto}, G. and {Contursi}, G. and {Cooper}, W.~J. and {Cornez}, T. and {Cowell}, S. and {Crifo}, F. and {Cropper}, M. and {Crosta}, M. and {Crowley}, C. and {Dafonte}, C. and {Dapergolas}, A. and {David}, M. and {David}, P. and {de Laverny}, P. and {De Luise}, F. and {De March}, R.},
        title = "{Gaia Data Release 3. Summary of the content and survey properties}",
      journal = {\aap},
         year = 2023,
        month = jun,
       volume = {674},
          eid = {A1},
        pages = {A1},
          doi = {10.1051/0004-6361/202243940},
archivePrefix = {arXiv},
       eprint = {2208.00211},
 primaryClass = {astro-ph.GA},
       adsurl = {https://ui.adsabs.harvard.edu/abs/2023A&A...674A...1G}
}

@ARTICLE{Isella2016,
       author = {{Isella}, Andrea and {Guidi}, Greta and {Testi}, Leonardo and {Liu}, Shangfei and {Li}, Hui and {Li}, Shengtai and {Weaver}, Erik and {Boehler}, Yann and {Carperter}, John M. and {De Gregorio-Monsalvo}, Itziar and {Manara}, Carlo F. and {Natta}, Antonella and {P{\'e}rez}, Laura M. and {Ricci}, Luca and {Sargent}, Anneila and {Tazzari}, Marco and {Turner}, Neal},
        title = "{Ringed Structures of the HD 163296 Protoplanetary Disk Revealed by ALMA}",
      journal = {\prl},
         year = 2016,
        month = dec,
       volume = {117},
       number = {25},
          eid = {251101},
        pages = {251101},
          doi = {10.1103/PhysRevLett.117.251101},
       adsurl = {https://ui.adsabs.harvard.edu/abs/2016PhRvL.117y1101I}
}

@ARTICLE{Isella2018,
       author = {{Isella}, Andrea and {Huang}, Jane and {Andrews}, Sean M. and {Dullemond}, Cornelis P. and {Birnstiel}, Tilman and {Zhang}, Shangjia and {Zhu}, Zhaohuan and {Guzm{\'a}n}, Viviana V. and {P{\'e}rez}, Laura M. and {Bai}, Xue-Ning and {Benisty}, Myriam and {Carpenter}, John M. and {Ricci}, Luca and {Wilner}, David J.},
        title = "{The Disk Substructures at High Angular Resolution Project (DSHARP). IX. A High-definition Study of the HD 163296 Planet-forming Disk}",
      journal = {\apjl},
         year = 2018,
        month = dec,
       volume = {869},
       number = {2},
          eid = {L49},
        pages = {L49},
          doi = {10.3847/2041-8213/aaf747},
archivePrefix = {arXiv},
       eprint = {1812.04047},
 primaryClass = {astro-ph.SR},
       adsurl = {https://ui.adsabs.harvard.edu/abs/2018ApJ...869L..49I}
}

@ARTICLE{Oberg2021,
       author = {{{\"O}berg}, Karin I. and {Guzm{\'a}n}, Viviana V. and {Walsh}, Catherine and {Aikawa}, Yuri and {Bergin}, Edwin A. and {Law}, Charles J. and {Loomis}, Ryan A. and {Alarc{\'o}n}, Felipe and {Andrews}, Sean M. and {Bae}, Jaehan and {Bergner}, Jennifer B. and {Boehler}, Yann and {Booth}, Alice S. and {Bosman}, Arthur D. and {Calahan}, Jenny K. and {Cataldi}, Gianni and {Cleeves}, L. Ilsedore and {Czekala}, Ian and {Furuya}, Kenji and {Huang}, Jane and {Ilee}, John D. and {Kurtovic}, Nicolas T. and {Le Gal}, Romane and {Liu}, Yao and {Long}, Feng and {M{\'e}nard}, Fran{\c{c}}ois and {Nomura}, Hideko and {P{\'e}rez}, Laura M. and {Qi}, Chunhua and {Schwarz}, Kamber R. and {Sierra}, Anibal and {Teague}, Richard and {Tsukagoshi}, Takashi and {Yamato}, Yoshihide and {van't Hoff}, Merel L.~R. and {Waggoner}, Abygail R. and {Wilner}, David J. and {Zhang}, Ke},
        title = "{Molecules with ALMA at Planet-forming Scales (MAPS). I. Program Overview and Highlights}",
      journal = {\apjs},
         year = 2021,
        month = nov,
       volume = {257},
       number = {1},
          eid = {1},
        pages = {1},
          doi = {10.3847/1538-4365/ac1432},
archivePrefix = {arXiv},
       eprint = {2109.06268},
 primaryClass = {astro-ph.EP},
       adsurl = {https://ui.adsabs.harvard.edu/abs/2021ApJS..257....1O}
}

@ARTICLE{Zhang2021,
       author = {{Zhang}, Ke and {Booth}, Alice S. and {Law}, Charles J. and {Bosman}, Arthur D. and {Schwarz}, Kamber R. and {Bergin}, Edwin A. and {{\"O}berg}, Karin I. and {Andrews}, Sean M. and {Guzm{\'a}n}, Viviana V. and {Walsh}, Catherine and {Qi}, Chunhua and {van't Hoff}, Merel L.~R. and {Long}, Feng and {Wilner}, David J. and {Huang}, Jane and {Czekala}, Ian and {Ilee}, John D. and {Cataldi}, Gianni and {Bergner}, Jennifer B. and {Aikawa}, Yuri and {Teague}, Richard and {Bae}, Jaehan and {Loomis}, Ryan A. and {Calahan}, Jenny K. and {Alarc{\'o}n}, Felipe and {M{\'e}nard}, Fran{\c{c}}ois and {Le Gal}, Romane and {Sierra}, Anibal and {Yamato}, Yoshihide and {Nomura}, Hideko and {Tsukagoshi}, Takashi and {P{\'e}rez}, Laura M. and {Trapman}, Leon and {Liu}, Yao and {Furuya}, Kenji},
        title = "{Molecules with ALMA at Planet-forming Scales (MAPS). V. CO Gas Distributions}",
      journal = {\apjs},
         year = 2021,
        month = nov,
       volume = {257},
       number = {1},
          eid = {5},
        pages = {5},
          doi = {10.3847/1538-4365/ac1580},
archivePrefix = {arXiv},
       eprint = {2109.06233},
 primaryClass = {astro-ph.EP},
       adsurl = {https://ui.adsabs.harvard.edu/abs/2021ApJS..257....5Z}
}

@ARTICLE{Pinte2018,
       author = {{Pinte}, C. and {Price}, D.~J. and {M{\'e}nard}, F. and {Duch{\^e}ne}, G. and {Dent}, W.~R.~F. and {Hill}, T. and {de Gregorio-Monsalvo}, I. and {Hales}, A. and {Mentiplay}, D.},
        title = "{Kinematic Evidence for an Embedded Protoplanet in a Circumstellar Disk}",
      journal = {\apjl},
         year = 2018,
        month = jun,
       volume = {860},
       number = {1},
          eid = {L13},
        pages = {L13},
          doi = {10.3847/2041-8213/aac6dc},
archivePrefix = {arXiv},
       eprint = {1805.10293},
 primaryClass = {astro-ph.SR},
       adsurl = {https://ui.adsabs.harvard.edu/abs/2018ApJ...860L..13P}
}

@ARTICLE{Izuquierdo2022,
       author = {{Izquierdo}, Andr{\'e}s F. and {Facchini}, Stefano and {Rosotti}, Giovanni P. and {van Dishoeck}, Ewine F. and {Testi}, Leonardo},
        title = "{A New Planet Candidate Detected in a Dust Gap of the Disk around HD 163296 through Localized Kinematic Signatures: An Observational Validation of the DISCMINER}",
      journal = {\apj},
         year = 2022,
        month = mar,
       volume = {928},
       number = {1},
          eid = {2},
        pages = {2},
          doi = {10.3847/1538-4357/ac474d},
archivePrefix = {arXiv},
       eprint = {2111.06367},
 primaryClass = {astro-ph.EP},
       adsurl = {https://ui.adsabs.harvard.edu/abs/2022ApJ...928....2I}
}

@ARTICLE{Calcino2022,
       author = {{Calcino}, Josh and {Hilder}, Thomas and {Price}, Daniel J. and {Pinte}, Christophe and {Bollati}, Francesco and {Lodato}, Giuseppe and {Norfolk}, Brodie J.},
        title = "{Mapping the Planetary Wake in HD 163296 with Kinematics}",
      journal = {\apjl},
         year = 2022,
        month = apr,
       volume = {929},
       number = {2},
          eid = {L25},
        pages = {L25},
          doi = {10.3847/2041-8213/ac64a7},
archivePrefix = {arXiv},
       eprint = {2111.07416},
 primaryClass = {astro-ph.EP},
       adsurl = {https://ui.adsabs.harvard.edu/abs/2022ApJ...929L..25C}
}

@ARTICLE{Fairlamb2015,
       author = {{Fairlamb}, J.~R. and {Oudmaijer}, R.~D. and {Mendigut{\'\i}a}, I. and {Ilee}, J.~D. and {van den Ancker}, M.~E.},
        title = "{A spectroscopic survey of Herbig Ae/Be stars with X-shooter - I. Stellar parameters and accretion rates}",
      journal = {\mnras},
         year = 2015,
        month = oct,
       volume = {453},
       number = {1},
        pages = {976-1001},
          doi = {10.1093/mnras/stv1576},
archivePrefix = {arXiv},
       eprint = {1507.05967},
 primaryClass = {astro-ph.SR},
       adsurl = {https://ui.adsabs.harvard.edu/abs/2015MNRAS.453..976F}
}

@ARTICLE{Calahan2021,
       author = {{Calahan}, Jenny K. and {Bergin}, Edwin A. and {Zhang}, Ke and {Schwarz}, Kamber R. and {{\"O}berg}, Karin I. and {Guzm{\'a}n}, Viviana V. and {Walsh}, Catherine and {Aikawa}, Yuri and {Alarc{\'o}n}, Felipe and {Andrews}, Sean M. and {Bae}, Jaehan and {Bergner}, Jennifer B. and {Booth}, Alice S. and {Bosman}, Arthur D. and {Cataldi}, Gianni and {Czekala}, Ian and {Huang}, Jane and {Ilee}, John D. and {Law}, Charles J. and {Le Gal}, Romane and {Long}, Feng and {Loomis}, Ryan A. and {M{\'e}nard}, Fran{\c{c}}ois and {Nomura}, Hideko and {Qi}, Chunhua and {Teague}, Richard and {van't Hoff}, Merel L.~R. and {Wilner}, David J. and {Yamato}, Yoshihide},
        title = "{Molecules with ALMA at Planet-forming Scales (MAPS). XVII. Determining the 2D Thermal Structure of the HD 163296 Disk}",
      journal = {\apjs},
         year = 2021,
        month = nov,
       volume = {257},
       number = {1},
          eid = {17},
        pages = {17},
          doi = {10.3847/1538-4365/ac143f},
archivePrefix = {arXiv},
       eprint = {2109.06202},
 primaryClass = {astro-ph.EP},
       adsurl = {https://ui.adsabs.harvard.edu/abs/2021ApJS..257...17C}
}

@ARTICLE{Qi2013,
       author = {{Qi}, Chunhua and {{\"O}berg}, Karin I. and {Wilner}, David J. and {Rosenfeld}, Katherine A.},
        title = "{First Detection of c-C$_{3}$H$_{2}$ in a Circumstellar Disk}",
      journal = {\apjl},
         year = 2013,
        month = mar,
       volume = {765},
       number = {1},
          eid = {L14},
        pages = {L14},
          doi = {10.1088/2041-8205/765/1/L14},
archivePrefix = {arXiv},
       eprint = {1302.0251},
 primaryClass = {astro-ph.GA},
       adsurl = {https://ui.adsabs.harvard.edu/abs/2013ApJ...765L..14Q}
}

@ARTICLE{Carney2017,
       author = {{Carney}, M.~T. and {Hogerheijde}, M.~R. and {Loomis}, R.~A. and {Salinas}, V.~N. and {{\"O}berg}, K.~I. and {Qi}, C. and {Wilner}, D.~J.},
        title = "{Increased H$_{2}$CO production in the outer disk around HD 163296}",
      journal = {\aap},
         year = 2017,
        month = sep,
       volume = {605},
          eid = {A21},
        pages = {A21},
          doi = {10.1051/0004-6361/201629342},
archivePrefix = {arXiv},
       eprint = {1705.10188},
 primaryClass = {astro-ph.SR},
       adsurl = {https://ui.adsabs.harvard.edu/abs/2017A&A...605A..21C}
}

@ARTICLE{Booth2019,
       author = {{Booth}, Alice S. and {Walsh}, Catherine and {Ilee}, John D. and {Notsu}, Shota and {Qi}, Chunhua and {Nomura}, Hideko and {Akiyama}, Eiji},
        title = "{The First Detection of $^{13}$C$^{17}$O in a Protoplanetary Disk: A Robust Tracer of Disk Gas Mass}",
      journal = {\apjl},
         year = 2019,
        month = sep,
       volume = {882},
       number = {2},
          eid = {L31},
        pages = {L31},
          doi = {10.3847/2041-8213/ab3645},
archivePrefix = {arXiv},
       eprint = {1908.05045},
 primaryClass = {astro-ph.EP},
       adsurl = {https://ui.adsabs.harvard.edu/abs/2019ApJ...882L..31B}
}

@ARTICLE{Guzman2021,
       author = {{Guzm{\'a}n}, Viviana V. and {Bergner}, Jennifer B. and {Law}, Charles J. and {{\"O}berg}, Karin I. and {Walsh}, Catherine and {Cataldi}, Gianni and {Aikawa}, Yuri and {Bergin}, Edwin A. and {Czekala}, Ian and {Huang}, Jane and {Andrews}, Sean M. and {Loomis}, Ryan A. and {Zhang}, Ke and {Le Gal}, Romane and {Alarc{\'o}n}, Felipe and {Ilee}, John D. and {Teague}, Richard and {Cleeves}, L. Ilsedore and {Wilner}, David J. and {Long}, Feng and {Schwarz}, Kamber R. and {Bosman}, Arthur D. and {P{\'e}rez}, Laura M. and {M{\'e}nard}, Fran{\c{c}}ois and {Liu}, Yao},
        title = "{Molecules with ALMA at Planet-forming Scales (MAPS). VI. Distribution of the Small Organics HCN, C$_{2}$H, and H$_{2}$CO}",
      journal = {\apjs},
         year = 2021,
        month = nov,
       volume = {257},
       number = {1},
          eid = {6},
        pages = {6},
          doi = {10.3847/1538-4365/ac1440},
archivePrefix = {arXiv},
       eprint = {2109.06391},
 primaryClass = {astro-ph.EP},
       adsurl = {https://ui.adsabs.harvard.edu/abs/2021ApJS..257....6G}
}

@ARTICLE{Ilee2021,
       author = {{Ilee}, John D. and {Walsh}, Catherine and {Booth}, Alice S. and {Aikawa}, Yuri and {Andrews}, Sean M. and {Bae}, Jaehan and {Bergin}, Edwin A. and {Bergner}, Jennifer B. and {Bosman}, Arthur D. and {Cataldi}, Gianni and {Cleeves}, L. Ilsedore and {Czekala}, Ian and {Guzm{\'a}n}, Viviana V. and {Huang}, Jane and {Law}, Charles J. and {Le Gal}, Romane and {Loomis}, Ryan A. and {M{\'e}nard}, Fran{\c{c}}ois and {Nomura}, Hideko and {{\"O}berg}, Karin I. and {Qi}, Chunhua and {Schwarz}, Kamber R. and {Teague}, Richard and {Tsukagoshi}, Takashi and {Wilner}, David J. and {Yamato}, Yoshihide and {Zhang}, Ke},
        title = "{Molecules with ALMA at Planet-forming Scales (MAPS). IX. Distribution and Properties of the Large Organic Molecules HC$_{3}$N, CH$_{3}$CN, and c-C$_{3}$H$_{2}$}",
      journal = {\apjs},
         year = 2021,
        month = nov,
       volume = {257},
       number = {1},
          eid = {9},
        pages = {9},
          doi = {10.3847/1538-4365/ac1441},
archivePrefix = {arXiv},
       eprint = {2109.06319},
 primaryClass = {astro-ph.EP},
       adsurl = {https://ui.adsabs.harvard.edu/abs/2021ApJS..257....9I}
}

@ARTICLE{Collings2004,
       author = {{Collings}, Mark P. and {Anderson}, Mark A. and {Chen}, Rui and {Dever}, John W. and {Viti}, Serena and {Williams}, David A. and {McCoustra}, Martin R.~S.},
        title = "{A laboratory survey of the thermal desorption of astrophysically relevant molecules}",
      journal = {\mnras},
         year = 2004,
        month = nov,
       volume = {354},
       number = {4},
        pages = {1133-1140},
          doi = {10.1111/j.1365-2966.2004.08272.x},
       adsurl = {https://ui.adsabs.harvard.edu/abs/2004MNRAS.354.1133C}
}

@ARTICLE{Altwegg2022,
       author = {{Altwegg}, K. and {Combi}, M. and {Fuselier}, S.~A. and {H{\"a}nni}, N. and {De Keyser}, J. and {Mahjoub}, A. and {M{\"u}ller}, D.~R. and {Pestoni}, B. and {Rubin}, M. and {Wampfler}, S.~F.},
        title = "{Abundant ammonium hydrosulphide embedded in cometary dust grains}",
      journal = {\mnras},
         year = 2022,
        month = nov,
       volume = {516},
       number = {3},
        pages = {3900-3910},
          doi = {10.1093/mnras/stac2440},
archivePrefix = {arXiv},
       eprint = {2208.11396},
 primaryClass = {astro-ph.EP},
       adsurl = {https://ui.adsabs.harvard.edu/abs/2022MNRAS.516.3900A}
}

@ARTICLE{vanderMarel2021,
       author = {{van der Marel}, Nienke and {Booth}, Alice S. and {Leemker}, Margot and {van Dishoeck}, Ewine F. and {Ohashi}, Satoshi},
        title = "{A major asymmetric ice trap in a planet-forming disk. I. Formaldehyde and methanol}",
      journal = {\aap},
         year = 2021,
        month = jul,
       volume = {651},
          eid = {L5},
        pages = {L5},
          doi = {10.1051/0004-6361/202141051},
archivePrefix = {arXiv},
       eprint = {2104.08906},
 primaryClass = {astro-ph.EP},
       adsurl = {https://ui.adsabs.harvard.edu/abs/2021A&A...651L...5V}
}

@ARTICLE{Booth2021_HD100546,
       author = {{Booth}, Alice S. and {Walsh}, Catherine and {Terwisscha van Scheltinga}, Jeroen and {van Dishoeck}, Ewine F. and {Ilee}, John D. and {Hogerheijde}, Michiel R. and {Kama}, Mihkel and {Nomura}, Hideko},
        title = "{An inherited complex organic molecule reservoir in a warm planet-hosting disk}",
      journal = {Nature Astronomy},
         year = 2021,
        month = jan,
       volume = {5},
        pages = {684-690},
          doi = {10.1038/s41550-021-01352-w},
archivePrefix = {arXiv},
       eprint = {2104.08348},
 primaryClass = {astro-ph.EP},
       adsurl = {https://ui.adsabs.harvard.edu/abs/2021NatAs...5..684B}
}

@ARTICLE{Brunken2022,
       author = {{Brunken}, Nashanty G.~C. and {Booth}, Alice S. and {Leemker}, Margot and {Nazari}, Pooneh and {van der Marel}, Nienke and {van Dishoeck}, Ewine F.},
        title = "{A major asymmetric ice trap in a planet-forming disk. III. First detection of dimethyl ether}",
      journal = {\aap},
         year = 2022,
        month = mar,
       volume = {659},
          eid = {A29},
        pages = {A29},
          doi = {10.1051/0004-6361/202142981},
archivePrefix = {arXiv},
       eprint = {2203.02936},
 primaryClass = {astro-ph.EP},
       adsurl = {https://ui.adsabs.harvard.edu/abs/2022A&A...659A..29B}
}

@ARTICLE{CASA,
       author = {{CASA Team} and {Bean}, Ben and {Bhatnagar}, Sanjay and {Castro}, Sandra and {Donovan Meyer}, Jennifer and {Emonts}, Bjorn and {Garcia}, Enrique and {Garwood}, Robert and {Golap}, Kumar and {Gonzalez Villalba}, Justo and {Harris}, Pamela and {Hayashi}, Yohei and {Hoskins}, Josh and {Hsieh}, Mingyu and {Jagannathan}, Preshanth and {Kawasaki}, Wataru and {Keimpema}, Aard and {Kettenis}, Mark and {Lopez}, Jorge and {Marvil}, Joshua and {Masters}, Joseph and {McNichols}, Andrew and {Mehringer}, David and {Miel}, Renaud and {Moellenbrock}, George and {Montesino}, Federico and {Nakazato}, Takeshi and {Ott}, Juergen and {Petry}, Dirk and {Pokorny}, Martin and {Raba}, Ryan and {Rau}, Urvashi and {Schiebel}, Darrell and {Schweighart}, Neal and {Sekhar}, Srikrishna and {Shimada}, Kazuhiko and {Small}, Des and {Steeb}, Jan-Willem and {Sugimoto}, Kanako and {Suoranta}, Ville and {Tsutsumi}, Takahiro and {van Bemmel}, Ilse M. and {Verkouter}, Marjolein and {Wells}, Akeem and {Xiong}, Wei and {Szomoru}, Arpad and {Griffith}, Morgan and {Glendenning}, Brian and {Kern}, Jeff},
        title = "{CASA, the Common Astronomy Software Applications for Radio Astronomy}",
      journal = {\pasp},
         year = 2022,
        month = nov,
       volume = {134},
       number = {1041},
          eid = {114501},
        pages = {114501},
          doi = {10.1088/1538-3873/ac9642},
archivePrefix = {arXiv},
       eprint = {2210.02276},
 primaryClass = {astro-ph.IM},
       adsurl = {https://ui.adsabs.harvard.edu/abs/2022PASP..134k4501C}
}

@ARTICLE{Booth2024,
       author = {{Booth}, Alice S. and {Drozdovskaya}, Maria N. and {Temmink}, Milou and {Nomura}, Hideko and {van Dishoeck}, Ewine F. and {Keyte}, Luke and {Law}, Charles J. and {Leemker}, Margot and {van der Marel}, Nienke and {Notsu}, Shota and {{\"O}berg}, Karin and {Walsh}, Catherine},
        title = "{Measuring the $^{34}$S and $^{33}$S Isotopic Ratios of Volatile Sulfur during Planet Formation}",
      journal = {\apj},
         year = 2024,
        month = nov,
       volume = {975},
       number = {1},
          eid = {72},
        pages = {72},
          doi = {10.3847/1538-4357/ad7817},
archivePrefix = {arXiv},
       eprint = {2409.03885},
 primaryClass = {astro-ph.EP},
       adsurl = {https://ui.adsabs.harvard.edu/abs/2024ApJ...975...72B}
}

@ARTICLE{Temmink2025,
       author = {{Temmink}, Milou and {Booth}, Alice S. and {Leemker}, Margot and {van der Marel}, Nienke and {van Dishoeck}, Ewine F. and {Evans}, Lucy and {Keyte}, Luke and {Law}, Charles J. and {Notsu}, Shota and {{\"O}berg}, Karin and {Walsh}, Catherine},
        title = "{Characterising the molecular line emission in the asymmetric Oph-IRS 48 dust trap: Temperatures, timescales, and sub-thermal excitation}",
      journal = {\aap},
         year = 2025,
        month = jan,
       volume = {693},
          eid = {A101},
        pages = {A101},
          doi = {10.1051/0004-6361/202452175},
archivePrefix = {arXiv},
       eprint = {2411.12418},
 primaryClass = {astro-ph.EP},
       adsurl = {https://ui.adsabs.harvard.edu/abs/2025A&A...693A.101T}
}

@ARTICLE{Fuente2023,
       author = {{Fuente}, A. and {Rivi{\`e}re-Marichalar}, P. and {Beitia-Antero}, L. and {Caselli}, P. and {Wakelam}, V. and {Esplugues}, G. and {Rodr{\'\i}guez-Baras}, M. and {Navarro-Almaida}, D. and {Gerin}, M. and {Kramer}, C. and {Bachiller}, R. and {Goicoechea}, J.~R. and {Jim{\'e}nez-Serra}, I. and {Loison}, J.~C. and {Ivlev}, A. and {Mart{\'\i}n-Dom{\'e}nech}, R. and {Spezzano}, S. and {Roncero}, O. and {Mu{\~n}oz-Caro}, G. and {Cazaux}, S. and {Marcelino}, N.},
        title = "{Gas phase Elemental abundances in Molecular cloudS (GEMS). VII. Sulfur elemental abundance}",
      journal = {\aap},
         year = 2023,
        month = feb,
       volume = {670},
          eid = {A114},
        pages = {A114},
          doi = {10.1051/0004-6361/202244843},
archivePrefix = {arXiv},
       eprint = {2212.03742},
 primaryClass = {astro-ph.GA},
       adsurl = {https://ui.adsabs.harvard.edu/abs/2023A&A...670A.114F}
}

@ARTICLE{Booth2025,
       author = {{Booth}, Alice S. and {Calahan}, Jenny and {Temmink}, Milou and {W{\"o}lfer}, Lisa and {Pegues}, Jamila and {Law}, Charles J. and {Evans}, Lucy and {Leemker}, Margot and {Notsu}, Shota and {{\"O}berg}, Karin and {Walsh}, Catherine and {van Dishoeck}, Ewine F.},
        title = "{The chemical diversity of giant-planet nurseries as revealed by ALMA}",
      journal = {arXiv e-prints},
         year = 2025,
        month = dec,
          eid = {arXiv:2512.01731},
        pages = {arXiv:2512.01731},
          doi = {10.48550/arXiv.2512.01731},
archivePrefix = {arXiv},
       eprint = {2512.01731},
 primaryClass = {astro-ph.EP},
       adsurl = {https://ui.adsabs.harvard.edu/abs/2025arXiv251201731B}
}

@ARTICLE{Keller2002,
       author = {{Keller}, L.~P. and {Hony}, S. and {Bradley}, J.~P. and {Molster}, F.~J. and {Waters}, L.~B.~F.~M. and {Bouwman}, J. and {de Koter}, A. and {Brownlee}, D.~E. and {Flynn}, G.~J. and {Henning}, T. and {Mutschke}, H.},
        title = "{Identification of iron sulphide grains in protoplanetary disks}",
      journal = {\nat},
         year = 2002,
        month = may,
       volume = {417},
       number = {6884},
        pages = {148-150},
          doi = {10.1038/417148a},
       adsurl = {https://ui.adsabs.harvard.edu/abs/2002Natur.417..148K}
}

@ARTICLE{Law2025_CSsurvey,
       author = {{Law}, Charles J. and {Le Gal}, Romane and {{\"O}berg}, Karin I. and {Zhang}, Ke and {Aikawa}, Yuri and {Andrews}, Sean M. and {Bae}, Jaehan and {Booth}, Alice S. and {Cataldi}, Gianni and {Cleeves}, L. Ilsedore and {Long}, Feng and {M{\'e}nard}, Fran{\c{c}}ois and {Qi}, Chunhua and {Teague}, Richard and {Wilner}, David J.},
        title = "{A Submillimeter Survey of CS Excitation in Protoplanetary Disks: Evidence of X-ray-Driven Sulfur Chemistry}",
      journal = {arXiv e-prints},
         year = 2025,
        month = nov,
          eid = {arXiv:2511.09628},
        pages = {arXiv:2511.09628},
          doi = {10.48550/arXiv.2511.09628},
archivePrefix = {arXiv},
       eprint = {2511.09628},
 primaryClass = {astro-ph.EP},
       adsurl = {https://ui.adsabs.harvard.edu/abs/2025arXiv251109628L}
}

@ARTICLE{Loomis2020,
       author = {{Loomis}, Ryan A. and {{\"O}berg}, Karin I. and {Andrews}, Sean M. and {Bergin}, Edwin and {Bergner}, Jennifer and {Blake}, Geoffrey A. and {Cleeves}, L. Ilsedore and {Czekala}, Ian and {Huang}, Jane and {Le Gal}, Romane and {M{\'e}nard}, Francois and {Pegues}, Jamila and {Qi}, Chunhua and {Walsh}, Catherine and {Williams}, Jonathan P. and {Wilner}, David J.},
        title = "{An Unbiased ALMA Spectral Survey of the LkCa 15 and MWC 480 Protoplanetary Disks}",
      journal = {\apj},
         year = 2020,
        month = apr,
       volume = {893},
       number = {2},
          eid = {101},
        pages = {101},
          doi = {10.3847/1538-4357/ab7cc8},
archivePrefix = {arXiv},
       eprint = {2006.16187},
 primaryClass = {astro-ph.SR},
       adsurl = {https://ui.adsabs.harvard.edu/abs/2020ApJ...893..101L}
}

@ARTICLE{Teague2025,
       author = {{Teague}, Richard and {Benisty}, Myriam and {Facchini}, Stefano and {Fukagawa}, Misato and {Pinte}, Christophe and {Andrews}, Sean M. and {Bae}, Jaehan and {Barraza-Alfaro}, Marcelo and {Cataldi}, Gianni and {Cuello}, Nicol{\'a}s and {Curone}, Pietro and {Czekala}, Ian and {Fasano}, Daniele and {Flock}, Mario and {Galloway-Sprietsma}, Maria and {Garg}, Himanshi and {Hall}, Cassandra and {Hammond}, Iain and {Hilder}, Thomas and {Huang}, Jane and {Ilee}, John D. and {Izquierdo}, Andr{\'e}s F. and {Kanagawa}, Kazuhiro and {Lesur}, Geoffroy and {Lodato}, Giuseppe and {Longarini}, Cristiano and {Loomis}, Ryan A. and {Masset}, Fr{\'e}d{\'e}ric and {Menard}, Francois and {Orihara}, Ryuta and {Price}, Daniel J. and {Rosotti}, Giovanni and {Stadler}, Jochen and {Testi}, Leonardo and {Yen}, Hsi-Wei and {Wafflard-Fernandez}, Gaylor and {Wilner}, David J. and {Winter}, Andrew J. and {W{\"o}lfer}, Lisa and {Yoshida}, Tomohiro C. and {Zawadzki}, Brianna},
        title = "{exoALMA. I. Science Goals, Project Design, and Data Products}",
      journal = {\apjl},
         year = 2025,
        month = may,
       volume = {984},
       number = {1},
          eid = {L6},
        pages = {L6},
          doi = {10.3847/2041-8213/adc43b},
archivePrefix = {arXiv},
       eprint = {2504.18688},
 primaryClass = {astro-ph.EP},
       adsurl = {https://ui.adsabs.harvard.edu/abs/2025ApJ...984L...6T}
}

@ARTICLE{Qi2015,
       author = {{Qi}, Chunhua and {{\"O}berg}, Karin I. and {Andrews}, Sean M. and {Wilner}, David J. and {Bergin}, Edwin A. and {Hughes}, A. Meredith and {Hogherheijde}, Michiel and {D'Alessio}, Paola},
        title = "{Chemical Imaging of the CO Snow Line in the HD 163296 Disk}",
      journal = {\apj},
         year = 2015,
        month = nov,
       volume = {813},
       number = {2},
          eid = {128},
        pages = {128},
          doi = {10.1088/0004-637X/813/2/128},
archivePrefix = {arXiv},
       eprint = {1510.00968},
 primaryClass = {astro-ph.SR},
       adsurl = {https://ui.adsabs.harvard.edu/abs/2015ApJ...813..128Q}
}

@ARTICLE{Law2021,
       author = {{Law}, Charles J. and {Loomis}, Ryan A. and {Teague}, Richard and {{\"O}berg}, Karin I. and {Czekala}, Ian and {Andrews}, Sean M. and {Huang}, Jane and {Aikawa}, Yuri and {Alarc{\'o}n}, Felipe and {Bae}, Jaehan and {Bergin}, Edwin A. and {Bergner}, Jennifer B. and {Boehler}, Yann and {Booth}, Alice S. and {Bosman}, Arthur D. and {Calahan}, Jenny K. and {Cataldi}, Gianni and {Cleeves}, L. Ilsedore and {Furuya}, Kenji and {Guzm{\'a}n}, Viviana V. and {Ilee}, John D. and {Le Gal}, Romane and {Liu}, Yao and {Long}, Feng and {M{\'e}nard}, Fran{\c{c}}ois and {Nomura}, Hideko and {Qi}, Chunhua and {Schwarz}, Kamber R. and {Sierra}, Anibal and {Tsukagoshi}, Takashi and {Yamato}, Yoshihide and {van't Hoff}, Merel L.~R. and {Walsh}, Catherine and {Wilner}, David J. and {Zhang}, Ke},
        title = "{Molecules with ALMA at Planet-forming Scales (MAPS). III. Characteristics of Radial Chemical Substructures}",
      journal = {\apjs},
         year = 2021,
        month = nov,
       volume = {257},
       number = {1},
          eid = {3},
        pages = {3},
          doi = {10.3847/1538-4365/ac1434},
archivePrefix = {arXiv},
       eprint = {2109.06210},
 primaryClass = {astro-ph.EP},
       adsurl = {https://ui.adsabs.harvard.edu/abs/2021ApJS..257....3L}
}

@ARTICLE{Hernandez-Vera2024,
       author = {{Hern{\'a}ndez-Vera}, Claudio and {Guzm{\'a}n}, Viviana V. and {Artur de la Villarmois}, Elizabeth and {{\"O}berg}, Karin I. and {Cleeves}, L. Ilsedore and {Hogerheijde}, Michiel R. and {Qi}, Chunhua and {Carpenter}, John and {Fayolle}, Edith C.},
        title = "{Radial and Vertical Constraints on the Icy Origin of H$_{2}$CO in the HD 163296 Protoplanetary Disk}",
      journal = {\apj},
         year = 2024,
        month = may,
       volume = {967},
       number = {1},
          eid = {68},
        pages = {68},
          doi = {10.3847/1538-4357/ad3cdb},
archivePrefix = {arXiv},
       eprint = {2404.06133},
 primaryClass = {astro-ph.EP},
       adsurl = {https://ui.adsabs.harvard.edu/abs/2024ApJ...967...68H}
}

@ARTICLE{Kashyap2025,
       author = {{Kashyap}, Parashmoni and {Majumdar}, Liton and {Bergin}, Edwin A. and {Blake}, Geoffrey A. and {Willacy}, Karen and {Guilloteau}, St{\'e}phane and {Dutrey}, Anne and {Liu}, Sheng-Yuan and {Henning}, Thomas and {Goldsmith}, Paul F. and {Lis}, Dariusz C. and {Maitrey}, S. and {Turner}, Neal and {Sahai}, Raghvendra and {Lee}, Chin-Fei and {Saito}, Masao},
        title = "{Unveiling the Chemical Complexity and C/O Ratio of the HD 163296 Protoplanetary Disk: Constraints from Multi-line ALMA Observations of Organics, Nitriles, Sulfur-bearing, and Deuterated Molecules}",
      journal = {arXiv e-prints},
         year = 2025,
        month = nov,
          eid = {arXiv:2511.10882},
        pages = {arXiv:2511.10882},
          doi = {10.48550/arXiv.2511.10882},
archivePrefix = {arXiv},
       eprint = {2511.10882},
 primaryClass = {astro-ph.EP},
       adsurl = {https://ui.adsabs.harvard.edu/abs/2025arXiv251110882K}
}

@ARTICLE{Yamato2023,
       author = {{Yamato}, Yoshihide and {Aikawa}, Yuri and {Ohashi}, Nagayoshi and {Tobin}, John J. and {J{\o}rgensen}, Jes K. and {Takakuwa}, Shigehisa and {Aso}, Yusuke and {Sai}, Jinshi (Insa Choi) and {Flores}, Christian and {de Gregorio-Monsalvo}, Itziar and {Hirano}, Shingo and {Han}, Ilseung and {Kido}, Miyu and {Koch}, Patrick M. and {Kwon}, Woojin and {Lai}, Shih-Ping and {Lee}, Chang Won and {Lee}, Jeong-Eun and {Li}, Zhi-Yun and {Lin}, Zhe-Yu Daniel and {Looney}, Leslie W. and {Mori}, Shoji and {Narayanan}, Suchitra and {Phuong}, Nguyen Thi and {Saigo}, Kazuya and {Santamar{\'\i}a-Miranda}, Alejandro and {Sharma}, Rajeeb and {Thieme}, Travis J. and {Tomida}, Kengo and {van't Hoff}, Merel L.~R. and {Yen}, Hsi-Wei},
        title = "{Early Planet Formation in Embedded Disks (eDisk). IV. The Ringed and Warped Structure of the Disk around the Class I Protostar L1489 IRS}",
      journal = {\apj},
         year = 2023,
        month = jul,
       volume = {951},
       number = {1},
          eid = {11},
        pages = {11},
          doi = {10.3847/1538-4357/accd71},
archivePrefix = {arXiv},
       eprint = {2306.15408},
 primaryClass = {astro-ph.EP},
       adsurl = {https://ui.adsabs.harvard.edu/abs/2023ApJ...951...11Y}
}

@ARTICLE{Zagaria2025,
       author = {{Zagaria}, Francesco and {Jiang}, Haochang and {Cataldi}, Gianni and {Facchini}, Stefano and {Benisty}, Myriam and {Aikawa}, Yuri and {Andrews}, Sean and {Bae}, Jaehan and {Barraza-Alfaro}, Marcelo and {Curone}, Pietro and {Czekala}, Ian and {Fasano}, Daniele and {Hall}, Cassandra and {Hammond}, Iain and {Huang}, Jane and {Ilee}, John D. and {Izquierdo}, Andr{\'e}s F. and {Lawrence}, Jensen and {Lodato}, Giuseppe and {M{\'e}nard}, Fran{\c{c}}ois and {Pinte}, Christophe and {Rosotti}, Giovanni P. and {Stadler}, Jochen and {Teague}, Richard and {Testi}, Leonardo and {Wilner}, David and {Winter}, Andrew and {Yoshida}, Tomohiro},
        title = "{SO Emission in the Dynamically Perturbed Protoplanetary Disks around CQ Tau and MWC 758}",
      journal = {\apj},
         year = 2025,
        month = aug,
       volume = {989},
       number = {1},
          eid = {30},
        pages = {30},
          doi = {10.3847/1538-4357/ade683},
archivePrefix = {arXiv},
       eprint = {2506.16481},
 primaryClass = {astro-ph.EP},
       adsurl = {https://ui.adsabs.harvard.edu/abs/2025ApJ...989...30Z}
}

@ARTICLE{Yoshida2024,
       author = {{Yoshida}, Tomohiro C. and {Nomura}, Hideko and {Law}, Charles J. and {Teague}, Richard and {Shibaike}, Yuhito and {Furuya}, Kenji and {Tsukagoshi}, Takashi},
        title = "{Outflow Driven by a Protoplanet Embedded in the TW Hya Disk}",
      journal = {\apjl},
         year = 2024,
        month = aug,
       volume = {971},
       number = {1},
          eid = {L15},
        pages = {L15},
          doi = {10.3847/2041-8213/ad654c},
archivePrefix = {arXiv},
       eprint = {2407.14395},
 primaryClass = {astro-ph.EP},
       adsurl = {https://ui.adsabs.harvard.edu/abs/2024ApJ...971L..15Y}
}

@ARTICLE{Huang2023,
       author = {{Huang}, Jane and {Bergin}, Edwin A. and {Bae}, Jaehan and {Benisty}, Myriam and {Andrews}, Sean M.},
        title = "{Molecular Mapping of DR Tau's Protoplanetary Disk, Envelope, Outflow, and Large-scale Spiral Arm}",
      journal = {\apj},
         year = 2023,
        month = feb,
       volume = {943},
       number = {2},
          eid = {107},
        pages = {107},
          doi = {10.3847/1538-4357/aca89c},
archivePrefix = {arXiv},
       eprint = {2301.02674},
 primaryClass = {astro-ph.SR},
       adsurl = {https://ui.adsabs.harvard.edu/abs/2023ApJ...943..107H}
}

@ARTICLE{Law2023,
       author = {{Law}, Charles J. and {Booth}, Alice S. and {{\"O}berg}, Karin I.},
        title = "{SO and SiS Emission Tracing an Embedded Planet and Compact $^{12}$CO and $^{13}$CO Counterparts in the HD 169142 Disk}",
      journal = {\apjl},
         year = 2023,
        month = jul,
       volume = {952},
       number = {1},
          eid = {L19},
        pages = {L19},
          doi = {10.3847/2041-8213/acdfd0},
archivePrefix = {arXiv},
       eprint = {2306.13710},
 primaryClass = {astro-ph.EP},
       adsurl = {https://ui.adsabs.harvard.edu/abs/2023ApJ...952L..19L}
}

@ARTICLE{Dutrey2024,
       author = {{Dutrey}, A. and {Chapillon}, E. and {Guilloteau}, S. and {Tang}, Y.~W. and {Boccaletti}, A. and {Bouscasse}, L. and {Collin-Dufresne}, T. and {Di Folco}, E. and {Fuente}, A. and {Pi{\'e}tu}, V. and {Rivi{\`e}re-Marichalar}, P. and {Semenov}, D.},
        title = "{Sulfur monoxide (SO) as a shock tracer in protoplanetary disks: Case of AB Aurigae}",
      journal = {\aap},
         year = 2024,
        month = sep,
       volume = {689},
          eid = {L7},
        pages = {L7},
          doi = {10.1051/0004-6361/202451299},
archivePrefix = {arXiv},
       eprint = {2408.14276},
 primaryClass = {astro-ph.EP},
       adsurl = {https://ui.adsabs.harvard.edu/abs/2024A&A...689L...7D}
}

@ARTICLE{CDMS1,
       author = {{M{\"u}ller}, H.~S.~P. and {Thorwirth}, S. and {Roth}, D.~A. and {Winnewisser}, G.},
        title = "{The Cologne Database for Molecular Spectroscopy, CDMS}",
      journal = {\aap},
         year = 2001,
        month = apr,
       volume = {370},
        pages = {L49-L52},
          doi = {10.1051/0004-6361:20010367},
       adsurl = {https://ui.adsabs.harvard.edu/abs/2001A&A...370L..49M}
}

@ARTICLE{CDMS2,
       author = {{M{\"u}ller}, Holger S.~P. and {Schl{\"o}der}, Frank and {Stutzki}, J{\"u}rgen and {Winnewisser}, Gisbert},
        title = "{The Cologne Database for Molecular Spectroscopy, CDMS: a useful tool for astronomers and spectroscopists}",
      journal = {Journal of Molecular Structure},
         year = 2005,
        month = may,
       volume = {742},
       number = {1-3},
        pages = {215-227},
          doi = {10.1016/j.molstruc.2005.01.027},
       adsurl = {https://ui.adsabs.harvard.edu/abs/2005JMoSt.742..215M}
}

@ARTICLE{CDMS3,
       author = {{Endres}, Christian P. and {Schlemmer}, Stephan and {Schilke}, Peter and {Stutzki}, J{\"u}rgen and {M{\"u}ller}, Holger S.~P.},
        title = "{The Cologne Database for Molecular Spectroscopy, CDMS, in the Virtual Atomic and Molecular Data Centre, VAMDC}",
      journal = {Journal of Molecular Spectroscopy},
         year = 2016,
        month = sep,
       volume = {327},
        pages = {95-104},
          doi = {10.1016/j.jms.2016.03.005},
       adsurl = {https://ui.adsabs.harvard.edu/abs/2016JMoSp.327...95E}
}

@ARTICLE{Helminger1972,
       author = {{Helminger}, Paul and {Cook}, Robert L. and {De Lucia}, Frank C.},
        title = "{Microwave Spectrum and Centrifugal Distortion Effects of H$_{2}$S}",
      journal = {\jcp},
         year = 1972,
        month = may,
       volume = {56},
       number = {9},
        pages = {4581-4584},
          doi = {10.1063/1.1677906},
       adsurl = {https://ui.adsabs.harvard.edu/abs/1972JChPh..56.4581H}
}

@ARTICLE{Belov1995,
       author = {{Belov}, S.~P. and {Yamada}, K.~M.~T. and {Winnewisser}, G. and {Poteau}, L. and {Bocquet}, R. and {Demaison}, J. and {Polyansky}, O. and {Tretyakov}, M.~Y.},
        title = "{Terahertz Rotational Spectrum of H $_{2}$S}",
      journal = {Journal of Molecular Spectroscopy},
         year = 1995,
        month = oct,
       volume = {173},
       number = {2},
        pages = {380-390},
          doi = {10.1006/jmsp.1995.1242},
       adsurl = {https://ui.adsabs.harvard.edu/abs/1995JMoSp.173..380B}
}

@ARTICLE{Klaus1996,
       author = {{Klaus}, Th. and {Saleck}, A.~H. and {Belov}, S.~P. and {Winnewisser}, G. and {Hirahara}, Y. and {Hayashi}, M. and {Kagi}, E. and {Kawaguchi}, K.},
        title = "{Pure Rotational Spectra of SO: Rare Isotopomers in the 80-GHz to 1.1-THz Region}",
      journal = {Journal of Molecular Spectroscopy},
         year = 1996,
        month = dec,
       volume = {180},
       number = {2},
        pages = {197-206},
          doi = {10.1006/jmsp.1996.0243},
       adsurl = {https://ui.adsabs.harvard.edu/abs/1996JMoSp.180..197K}
}

@ARTICLE{Clark1976,
       author = {{Clark}, William W. and {De Lucia}, Frank C.},
        title = "{The microwave spectrum and rotational structure of the $^{1}${\ensuremath{\Delta}} and $^{3}${\ensuremath{\Sigma}} electronic states of sulfur monoxide}",
      journal = {Journal of Molecular Spectroscopy},
         year = 1976,
        month = mar,
       volume = {60},
       number = {1-3},
        pages = {332-342},
          doi = {10.1016/0022-2852(76)90136-3},
       adsurl = {https://ui.adsabs.harvard.edu/abs/1976JMoSp..60..332C}
}

@ARTICLE{Muller2005,
       author = {{M{\"u}ller}, Holger S.~P. and {Br{\"u}nken}, Sandra},
        title = "{Accurate rotational spectroscopy of sulfur dioxide, SO$_{2}$, in its ground vibrational and first excited bending states, v$_{2}$ = 0, 1, up to 2 THz}",
      journal = {Journal of Molecular Spectroscopy},
         year = 2005,
        month = aug,
       volume = {232},
       number = {2},
        pages = {213-222},
          doi = {10.1016/j.jms.2005.04.010},
       adsurl = {https://ui.adsabs.harvard.edu/abs/2005JMoSp.232..213M}
}

@ARTICLE{Belov1998,
       author = {{Belov}, S.~P. and {Tretyakov}, M.~Y. and {Kozin}, I.~N. and {Klisch}, E. and {Winnewisser}, G. and {Lafferty}, W.~J. and {Flaud}, J.-M.},
        title = "{High Frequency Transitions in the Rotational Spectrum of SO $_{2}$}",
      journal = {Journal of Molecular Spectroscopy},
         year = 1998,
        month = sep,
       volume = {191},
       number = {1},
        pages = {17-27},
          doi = {10.1006/jmsp.1998.7576},
       adsurl = {https://ui.adsabs.harvard.edu/abs/1998JMoSp.191...17B}
}

@ARTICLE{Oba2018,
       author = {{Oba}, Y. and {Tomaru}, T. and {Lamberts}, T. and {Kouchi}, A. and {Watanabe}, N.},
        title = "{An infrared measurement of chemical desorption from interstellar ice analogues}",
      journal = {Nature Astronomy},
         year = 2018,
        month = mar,
       volume = {2},
        pages = {228-232},
          doi = {10.1038/s41550-018-0380-9},
archivePrefix = {arXiv},
       eprint = {1810.04669},
 primaryClass = {astro-ph.GA},
       adsurl = {https://ui.adsabs.harvard.edu/abs/2018NatAs...2..228O}
}

@ARTICLE{Fuente2017,
       author = {{Fuente}, Asunci{\'o}n and {Goicoechea}, Javier R. and {Pety}, J{\'e}r{\^o}me and {Le Gal}, Romane and {Mart{\'\i}n-Dom{\'e}nech}, Rafael and {Gratier}, Pierre and {Guzm{\'a}n}, Viviana and {Roueff}, Evelyne and {Loison}, Jean Christophe and {Mu{\~n}oz Caro}, Guillermo M. and {Wakelam}, Valentine and {Gerin}, Maryvonne and {Riviere-Marichalar}, Pablo and {Vidal}, Thomas},
        title = "{First Detection of Interstellar S$_{2}$H}",
      journal = {\apjl},
         year = 2017,
        month = dec,
       volume = {851},
       number = {2},
          eid = {L49},
        pages = {L49},
          doi = {10.3847/2041-8213/aaa01b},
archivePrefix = {arXiv},
       eprint = {1712.03036},
 primaryClass = {astro-ph.GA},
       adsurl = {https://ui.adsabs.harvard.edu/abs/2017ApJ...851L..49F}
}

@ARTICLE{Wassell2006,
       author = {{Wassell}, E.~J. and {Grady}, C.~A. and {Woodgate}, B. and {Kimble}, R.~A. and {Bruhweiler}, F.~C.},
        title = "{An Asymmetric Outflow from the Herbig Ae Star HD 163296}",
      journal = {\apj},
         year = 2006,
        month = oct,
       volume = {650},
       number = {2},
        pages = {985-997},
          doi = {10.1086/507268},
       adsurl = {https://ui.adsabs.harvard.edu/abs/2006ApJ...650..985W}
}

@ARTICLE{Ellerbroek2014,
       author = {{Ellerbroek}, L.~E. and {Podio}, L. and {Dougados}, C. and {Cabrit}, S. and {Sitko}, M.~L. and {Sana}, H. and {Kaper}, L. and {de Koter}, A. and {Klaassen}, P.~D. and {Mulders}, G.~D. and {Mendigut{\'\i}a}, I. and {Grady}, C.~A. and {Grankin}, K. and {van Winckel}, H. and {Bacciotti}, F. and {Russell}, R.~W. and {Lynch}, D.~K. and {Hammel}, H.~B. and {Beerman}, L.~C. and {Day}, A.~N. and {Huelsman}, D.~M. and {Werren}, C. and {Henden}, A. and {Grindlay}, J.},
        title = "{Relating jet structure to photometric variability: the Herbig Ae star HD 163296}",
      journal = {\aap},
         year = 2014,
        month = mar,
       volume = {563},
          eid = {A87},
        pages = {A87},
          doi = {10.1051/0004-6361/201323092},
archivePrefix = {arXiv},
       eprint = {1401.3744},
 primaryClass = {astro-ph.SR},
       adsurl = {https://ui.adsabs.harvard.edu/abs/2014A&A...563A..87E}
}

@ARTICLE{Booth2021_HD163296,
       author = {{Booth}, Alice S. and {Tabone}, Beno{\^\i}t and {Ilee}, John D. and {Walsh}, Catherine and {Aikawa}, Yuri and {Andrews}, Sean M. and {Bae}, Jaehan and {Bergin}, Edwin A. and {Bergner}, Jennifer B. and {Bosman}, Arthur D. and {Calahan}, Jenny K. and {Cataldi}, Gianni and {Cleeves}, L. Ilsedore and {Czekala}, Ian and {Guzm{\'a}n}, Viviana V. and {Huang}, Jane and {Law}, Charles J. and {Le Gal}, Romane and {Long}, Feng and {Loomis}, Ryan A. and {M{\'e}nard}, Fran{\c{c}}ois and {Nomura}, Hideko and {{\"O}berg}, Karin I. and {Qi}, Chunhua and {Schwarz}, Kamber R. and {Teague}, Richard and {Tsukagoshi}, Takashi and {Wilner}, David J. and {Yamato}, Yoshihide and {Zhang}, Ke},
        title = "{Molecules with ALMA at Planet-forming Scales (MAPS). XVI. Characterizing the Impact of the Molecular Wind on the Evolution of the HD 163296 System}",
      journal = {\apjs},
         year = 2021,
        month = nov,
       volume = {257},
       number = {1},
          eid = {16},
        pages = {16},
          doi = {10.3847/1538-4365/ac1ad4},
archivePrefix = {arXiv},
       eprint = {2109.06586},
 primaryClass = {astro-ph.EP},
       adsurl = {https://ui.adsabs.harvard.edu/abs/2021ApJS..257...16B}
}
\bibliographystyle{aasjournal}



\end{document}